\documentclass[journal]{IEEEtran}
\usepackage[noadjust]{cite}
\usepackage{srcltx}
\usepackage{graphicx}
\usepackage{epstopdf}
\usepackage{psfrag}
\usepackage{epsfig}
\usepackage[tbtags,sumlimits,nointlimits,reqno]{amsmath}
\usepackage{amssymb}
\usepackage{xcolor}
\usepackage{float}
\usepackage{subfigure}
\usepackage[most]{tcolorbox}
\usepackage{soul}
\usepackage{footnote}
\usepackage{amsmath}
\usepackage{color}
\newcommand{\jstyle}[1]{\textit{\textbf{{1}}}}
\newcommand{\rstyle}[1]{\textit{\textbf{{1}}}}
\newcommand{\cstyle}[1]{\textit{\textbf{{1}}}}
\newcommand{\bstyle}[1]{\textit{\textbf{{1}}}}
\newcommand{\istyle}[1]{\emph{\textbf{{1}}}}
\newcommand{\tbpstyle}[1]{\textit{{1}}}
\usepackage[framemethod=TikZ]{mdframed}
\usepackage{lipsum}
\mdfdefinestyle{MyFrame}{%
	linecolor=blue,
	outerlinewidth=2pt,
	roundcorner=20pt,
	innertopmargin=\baselineskip,
	innerbottommargin=\baselineskip,
	innerrightmargin=20pt,
	innerleftmargin=20pt,
	backgroundcolor=gray!50!white}

\begin{document}
\title{The Bragg Frequency Convertor: A Meeting Between Spatial and Temporal Periodicities For Selective Parametric Frequency Translation}
\author{Sajjad Taravati,~\IEEEmembership{Senior Member,~IEEE}
		\thanks{S. Taravati is with the School of Electronics and Computer Science, University of Southampton, Southampton SO17 1BJ, UK (e-mail: s.taravati@soton.ac.uk).}%
		\thanks{Manuscript received *, 2025; revised *, 2026.}
}
\markboth{IEEE Transaction on Antennas and Propagation,~Vol.~*, No.~*, *~2026}%
{* \MakeLowercase{\textit{et al.}}: Bare Demo of IEEEtran.cls for IEEE Journals}

\maketitle
	
\begin{abstract}
This study introduces the Bragg Frequency Converter, a spatiotemporal-periodic grating concept that extends conventional Bragg gratings into the dynamic domain for pure parametric frequency conversion. By selectively time-modulating either the high-index or low-index layers of a quarter-wave stack, the structure achieves directional frequency conversion: high-index modulation yields efficient down-conversion, while low-index modulation produces up-conversion. This layer selectivity stems from the asymmetric Bloch mode distribution and associated phase-matching conditions. One practical realization, based on a silicon rib waveguide with periodic sidewall corrugations and selective doping of the high-index segments, is presented and analyzed. A coupled-mode theory is developed to explain the mechanism and validated through full-wave simulations. An experimental setup using optical pumping is also proposed for practical implementation. The theoretical and numerical results establish temporal Bragg gratings as a versatile, reconfigurable platform for spurious-free frequency conversion with applications in optical signal processing and integrated photonics.
\end{abstract}
	
	
\IEEEpeerreviewmaketitle
	
\section{Introduction}\label{sec:introduction}
\IEEEPARstart{T}he pursuit of dynamic control over light-matter interaction has driven optics and photonics research beyond static structures and into the domain of dynamic media~\cite{Taravati_Kishk_MicMag_2019,sisler2024electrically,Taravati_NC_2021,taravati2024spatiotemporal,taravati20234d,taravati2025designing,taravati2025finite,taravati2025designing,patel2026photonic,chen2026novel,li2026tunable}. By judiciously modulating the electromagnetic properties of a material in time, it is possible to break Lorentz reciprocity, create non-reciprocal devices, and realize optical isolation without magnetic materials~\cite{Taravati_PRB_SB_2017,Taravati_AMTech_2021,koutzoglou2026robust}. More profoundly, temporal modulation can act as a synthetic pump that injects energy into an optical system, enabling parametric processes such as nonreciprocal transmission and isolation~\cite{Taravati_PRB_2017,Chamanara_PRB_2017,Taravati_PRAp_2018,taravati2020full,taravati2025_entangle,yilmaz2025numerical}, spatiotemporal diffraction~\cite{taravati_PRApp_2019}, amplification~\cite{cullen1958travelling,tien1958traveling,Taravati_Kishk_PRB_2018,taravati2021programmable,taravati2025temporal,horsley2024traveling}, frequency conversion~\cite{taravati2016mixer,taravati2021pure,Taravati_PRB_Mixer_2018,Taravati_ACSP_2022,taravati2024efficient,kovalev2025parametric}, beamsplitting~\cite{Taravati_Kishk_PRB_2018}, absorption~\cite{taravati2024one,taravati2024spatiotemporal}, multifunctionality~\cite{Taravati_AMA_PRApp_2020}, and wave mixing and frequency multiplexing~\cite{taravati2015space,Taravati_Kishk_TAP_2019,taravati2025light,sabri2021broadband,taravati2025frequency}. Among these parametric phenomena, pure frequency conversion~\cite{taravati2021pure,Taravati_PRB_Mixer_2018}, the coherent translation of an optical signal from one frequency to another with minimal residual power at the original frequency, stands as a cornerstone for communications, spectral imaging, and frequency-domain signal processing.

Conventional approaches to frequency conversion rely on nonlinear optical crystals or waveguides, where a strong pump beam mediates energy transfer between signal and idler waves via second- or third-order nonlinearities. While effective, these methods often require stringent conditions, high pump powers, and bulk materials, limiting their integration into compact photonic circuits. However, achieving pure conversion, where the output is dominated by the target frequency with strong suppression of the original carrier, remains challenging in simple modulated waveguides, as the input frequency typically co-propagates and transmits alongside the generated sidebands. In parallel, spatially periodic structures, photonic crystals and Bragg gratings, have long been used to engineer the density of optical states and control the propagation of light~\cite{othonos1997fiber,hill2002fiber}. A Bragg grating, in its simplest form, is a stack of alternating high- and low-index layers designed to create a photonic stopband: a range of frequencies over which light is strongly reflected~\cite{giles2002lightwave,burla2013integrated,albert2013tilted,khalil2022electrically,rohan2024recent,la2024bragg,xue2022real,chen2025high,bruckerhoff2025general,yeh1990optical}. However, a critical question remains: can we achieve pure frequency conversion, where the output is dominated by the shifted frequency with the original carrier suppressed, in a compact, integrated platform?

In this work, we introduce the Bragg Frequency Converter, a spatiotemporal grating concept that achieves pure parametric frequency conversion through the interplay of spatial and temporal periodicities. By selectively time-modulating either the high-index or low-index layers of a quarter-wave Bragg stack, we theoretically demonstrate layer-dependent conversion: high-index modulation yields efficient down-conversion, while low-index modulation produces up-conversion. This selectivity originates from the asymmetric Bloch mode distribution in the structure. The carrier is strongly reflected by the spatial Bragg periodicity, while the generated sidebands lie outside the stopband and propagate freely in both directions. Through coherent summation of distributed temporal scattering events, a single clean converted frequency dominates the output. We develop a coupled-mode theory explaining the layer-selective behavior and validate it with full-wave simulations. A practical implementation and an experimental setup is proposed for future experimental validation. The proposed device provides a compact, all-dielectric platform for reconfigurable frequency conversion controlled by the choice of modulated layer and modulation phase.

\begin{figure*}
	\begin{center}
\subfigure[]{\label{Fig:sch2}
	\includegraphics[width=1.5\columnwidth]{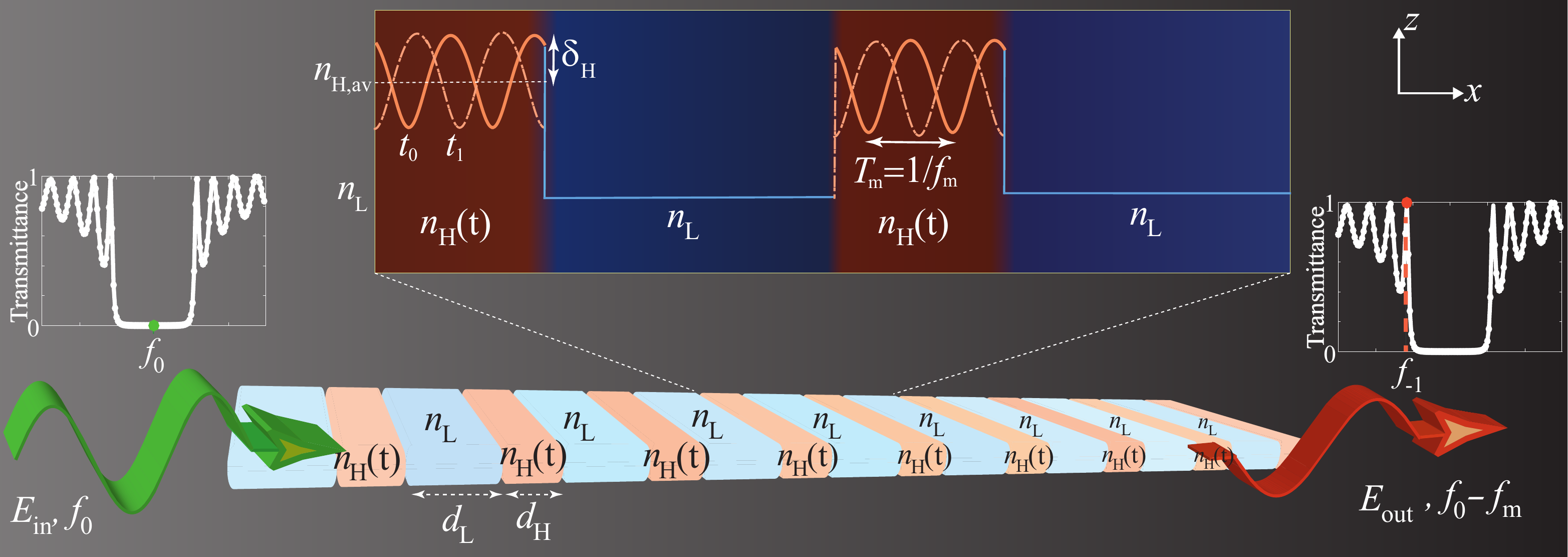}}
\subfigure[]{\label{Fig:sch1}
		\includegraphics[width=1.5\columnwidth]{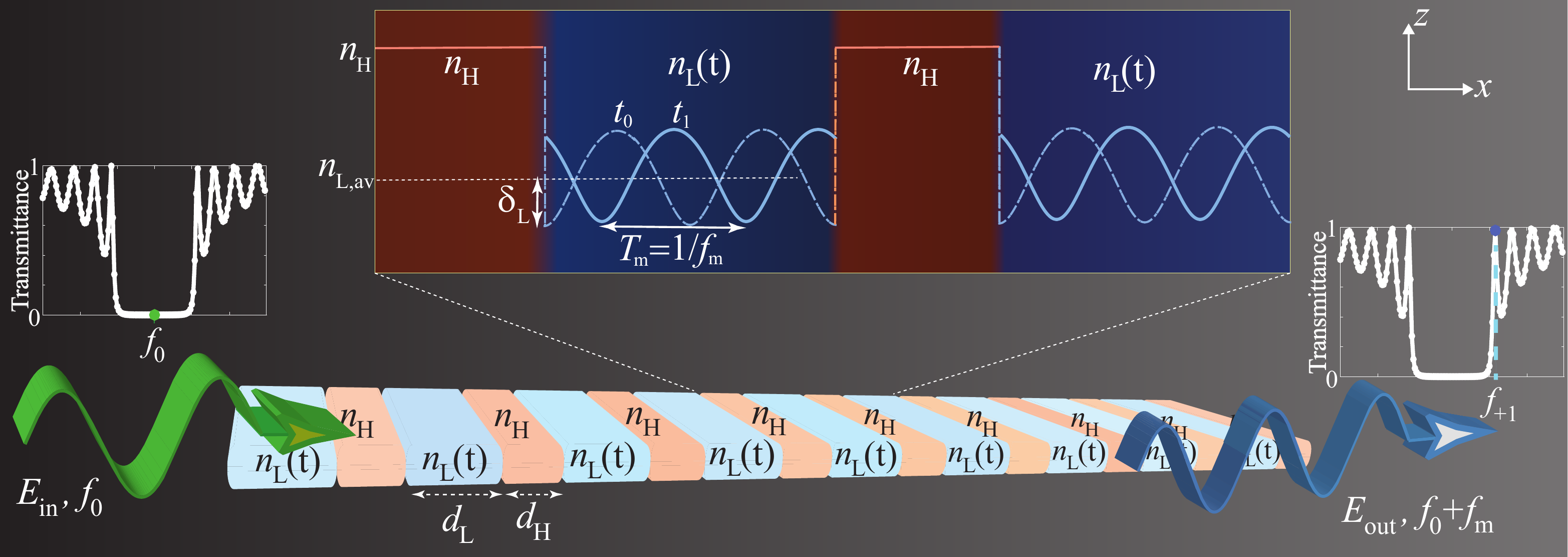}}
		\caption{Schematic of the Bragg Frequency Convertor. (a)~Down-conversion configuration, where high-index layers are time-modulated and the low-index layers remain static. This configuration preferentially couples to the lower band edge, enhancing down-conversion to \(f_{-1}\). The down-converted sideband lies outside the stopband and propagates freely, while the carrier \(f_0\) and other time harmonics suppressed by the Bragg stopband. (b)~Up-conversion configuration, where low-index layers are modulated targeting up-conversion to \(f_{+1} = f_0 + f_{\rm m} \), while the high-index layers remain static. Due to the Bloch mode structure of the quarter-wave stack, modulating the low-index layers preferentially couples to the upper band edge, enhancing up-conversion to \(f_{+1}\).}
		\label{Fig:sch}
	\end{center}
\end{figure*}

\section{Theory}
A Bragg grating is a periodic structure composed of alternating layers with high (\(n_\text{H}\)) and low (\(n_\text{L}\)) refractive indices, each having an optical thickness of \(\lambda_\text{B}/4\) at the design wavelength \(\lambda_\text{B}\). This quarter-wave condition ensures constructive interference of reflections from each interface, creating a photonic stopband centered at the Bragg frequency \(\omega_\text{B} = 2\pi c/\lambda_\text{B}\). When the refractive index of the Bragg grating layers is modulated in time, the system transforms from a passive filter into an active, time-variant medium capable of pure parametric frequency conversion in transmission. Consider the temporal Bragg grating in Fig.~\ref{Fig:sch}, where either the high-index or low-index layers experience sinusoidal temporal modulation
\begin{equation}\label{eq:refractive_index_time_a}
		n_\text{H/L}(t) = n_{\text{H/L},\text{av}} + \Delta n_\text{H/L} \cos(\omega_\text{m} t + \varphi_\text{H/L})
\end{equation}
where $\Delta n_\text{H}=\delta_\text{H} n_\text{H}$ and $\Delta n_\text{L}=\delta_\text{L}n_\text{L}$, in which $\delta_\text{H}$ and $\delta_\text{L}$ are the temporal modulation amplitudes of the high and low index layers, respectively. In addition, \(\omega_\text{m} = 2\pi f_\text{m}\) is the modulation frequency, \(\delta_{H}, \delta_{L} \ll n_{\text{av}}\) are the modulation depths, and \(\varphi_\text{H}, \varphi_\text{L}\) are modulation phases whose adjustment controls the amplitude of the converted output. The temporal Bragg grating consists of \(N\) periods, each with spatial period \(\Lambda\). The total physical length of the structure is $L = N \Lambda = N(d_\text{H} + d_\text{L})$, where \(d_\text{H} = \lambda_\text{B}/(4n_\text{H})\) and \(d_\text{L} = \lambda_\text{B}/(4n_\text{L})\) are the thicknesses of the high- and low-index layers, respectively. This length \(L\) determines the interaction distance for parametric processes and is critical for achieving efficient conversion. Crucially, the inherent Bragg reflection at the stopband center frequency \(\omega_0 \approx \omega_B\) suppresses transmission at the input frequency. This suppression, combined with parametric coupling, enables the generation of a pure converted signal at the sidebands. The temporal variation breaks time-translation symmetry, enabling energy transfer from the modulation pump at frequency $\omega_{\rm m}$ to the optical signal at $\omega_0$, producing output frequencies $\omega_{\text{out}} = \omega_0 \pm \omega_{\rm m}$ through parametric interaction. Remarkably, the direction of conversion is determined by the modulated layer type: modulation of the high-index layers (\(n_\text{H}\)) predominantly generates a down-converted signal at \(\omega_0 - \omega_\text{m}\), while modulation of the low-index layers (\(n_\text{L}\)) generates an up-converted signal at \(\omega_0 + \omega_\text{m}\).

\begin{figure}
	\begin{center}
		\includegraphics[width=1\columnwidth]{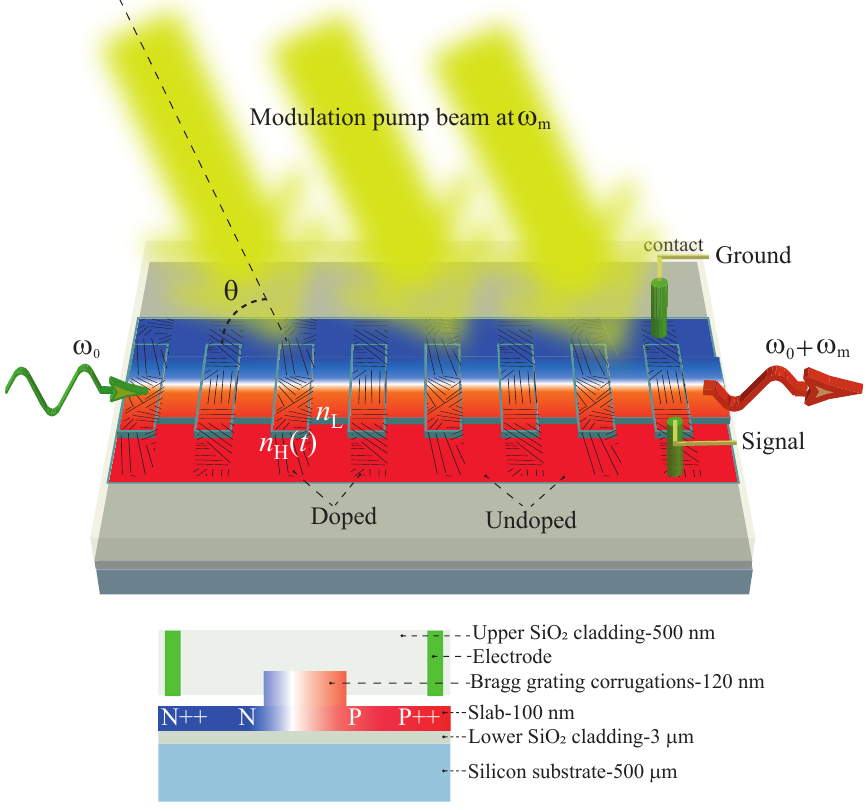}
		\caption{Schematic of the temporal Bragg grating frequency down-converter (N=8) in Fig.~\ref{Fig:sch2}. High-index (wider) corrugated segments are selectively doped to form lateral PN junctions for temporal modulation $  n_\text{H}(t)  $, while low-index (narrower) segments remain undoped. A common DC bias is applied to the doped regions. An obliquely incident optical pump beam at $  \omega_{\rm m} $ (incidence angle $  \theta  $) induces both time-varying index modulation and the desired phase gradient $  \Delta\phi  $.}
		\label{fig:fab}
	\end{center}
\end{figure}

Figure~\ref{fig:fab} demonstrates the structure of the proposed temporal Bragg grating frequency down-converter corresponding to operation in Fig.\ref{Fig:sch2}. The device is based on a silicon rib waveguide with periodic sidewall corrugations that create alternating high effective index ($  n_\text{H}  $, wider segments) and low effective index ($  n_\text{L}  $, narrower segments) regions. To achieve selective temporal modulation, lateral PN junctions are formed only in the high-index segments through selective doping (N++/N and P/P++ regions), while the low-index segments remain undoped and unmodulated. heavily doped N++ and P++ regions are placed in the thin slab areas on both sides of the rib to ensure low-resistance electrical contacts, while moderately doped N and P regions extend from the slab upward into the sides of the rib and along the corrugated sidewalls. A common DC bias is applied to the doped regions to optimize the average carrier density and minimize free-carrier absorption. The fast temporal refractive index modulation at frequency $  \omega_{\rm m}  $ is realized by an obliquely incident optical pump beam. The incidence angle $  \theta  $ is chosen to imprint the desired modulation phase gradient $  \Delta\phi  $ across the periods. For up-conversion operation in Fig.\ref{Fig:sch1}, the doping and modulation are applied to the low-index layers instead, leaving the high-index layers undoped.

\begin{figure}
	\begin{center}
		\includegraphics[width=1\columnwidth]{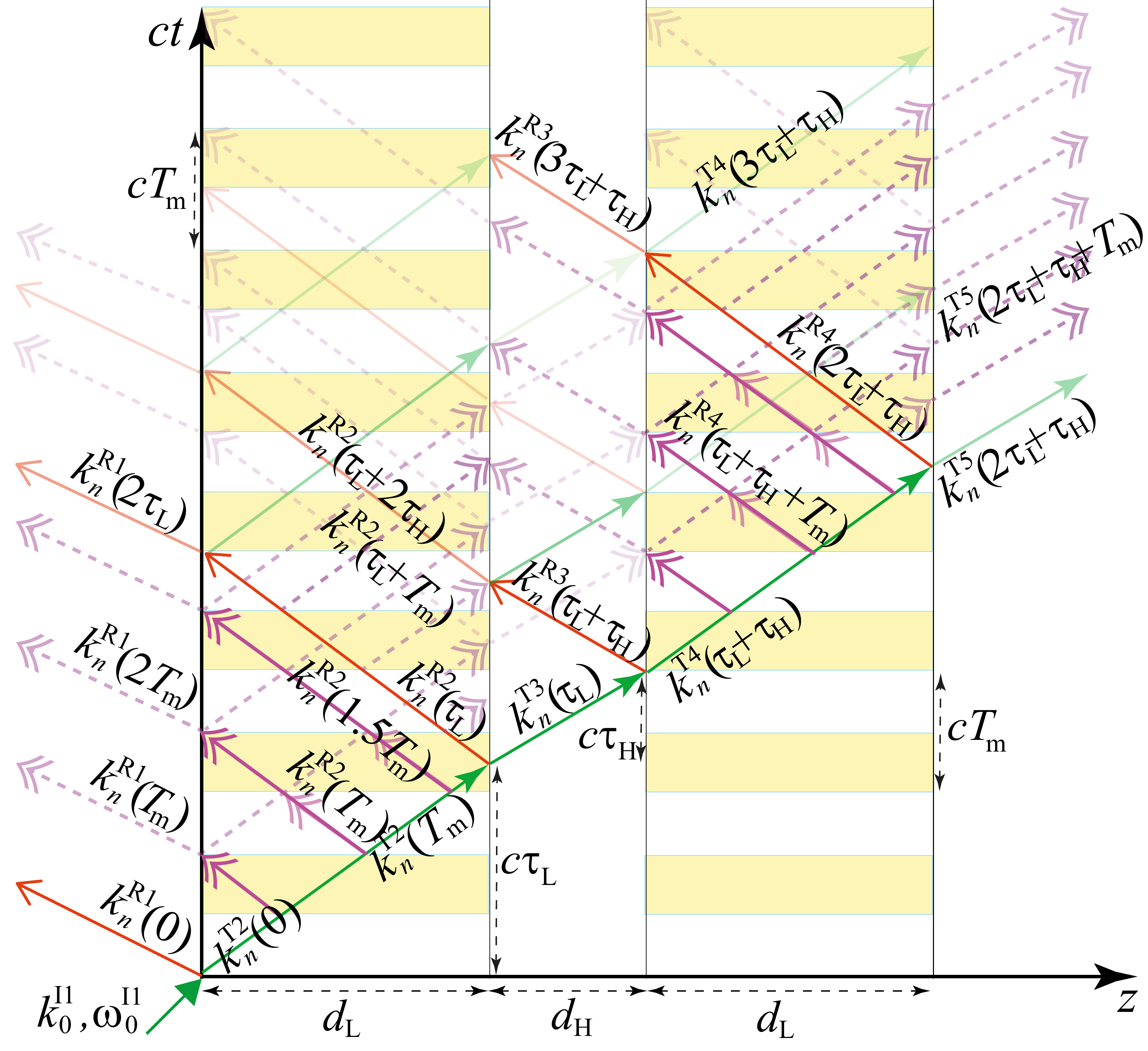}
		\caption{Space-time diagram of a temporal Bragg grating (up-conversion mode). where low-index layers are time-modulated (striped regions) and high-index layers are static (solid). Layer thicknesses \(d_\text{L}\) and \(d_\text{H}\) with corresponding transit times \(\tau_\text{L} = n_\text{L} d_\text{L}/c\) and \(\tau_\text{H} = n_\text{H} d_\text{H}/c\). Spatial period \(\Lambda = d_\text{L} + d_\text{H}\). Modulation period \(T_\text{m} = 2\pi/\omega_\text{m}\) indicated by vertical spacing \(cT_\text{m}\). Temporal scattering events recur with period \(T_\text{m}\) within each modulated layer.}
		\label{fig:stdiagram}
	\end{center}
\end{figure}

To visualize the wave propagation in a temporal Bragg grating, we construct a space-time diagram that tracks both spatial and temporal scattering events across multiple periods. Figure~\ref{fig:stdiagram} presents this diagram for the case of low-index layer modulation, revealing the intricate interplay between periodic spatial boundaries and periodic temporal modulation. The diagram uses spatial coordinate \(z\) on the horizontal axis and time \(ct\) on the vertical axis, with light lines indicating wave propagation trajectories. The structure consists of alternating low-index (\(n_\text{L}\), modulated) and high-index (\(n_\text{H}\), static) layers with physical thicknesses \(d_\text{L}\) and \(d_\text{H}\), respectively. The transit time through each layer reads
\begin{equation}
	\tau_\text{H/L} = \frac{d_\text{H/L}}{v_\text{H/L}} = \frac{n_\text{H/L} d_\text{H/L}}{c},
\end{equation}
where \(v_\text{L} = c/n_\text{L}\) and \(v_\text{H} = c/n_\text{H}\) are the group velocities (assuming negligible dispersion). These transit times determine the temporal spacing between successive scattering events within a single layer. Superimposed on these layer transit times is the modulation period \(T_\text{m} = 2\pi/\omega_\text{m}\), indicated by the vertical spacing \(cT_\text{m}\). The diagram spans multiple modulation periods, revealing how the dynamics repeat with this temporal periodicity while accumulating phase.

In Fig.~\ref{fig:stdiagram}, within each modulated low-index layer (striped regions), the time-varying index continuously generates new frequency components. These temporal scattering events—distributed throughout the layer volume, are shown as discrete instants for clarity. They include temporal reflections $k_n^\text{R1}(0)$, $k_n^\text{R2}(T_\text{m})$, $k_n^\text{R2}(T_\text{m}/2)$, $k_n^\text{R2}(3T_\text{m}/2), \ldots$ that are polychromatic backward-propagating waves generated within the modulated layer and propagating leftward, and temporal transmissions $k_n^\text{T2}(0)$, $k_n^\text{T2}(T_\text{m}/2)$, $k_n^\text{T2}(T_\text{m})$, $k_n^\text{T2}(3T_\text{m}/2), \ldots$ that are polychromatic forward-propagating waves generated within the modulated layer and propagating rightward. These temporal scattering events recur with period \(T_\text{m}\), forming a vertical array spaced by \(cT_\text{m}\) that visualizes the modulation periodicity. The frequency conversion outcome depends on which wave components interfere constructively at the output—a condition controlled by the phase \(\phi\) accumulated over multiple spatial and temporal periods. Tuning \(\phi\) maximizes the desired sideband. The directional asymmetry—up-conversion for low-index modulation, down-conversion for high-index modulation—originates from the Bloch mode profiles of the quarter-wave stack. At the lower band edge (just below the stopband), field peaks in high-index layers; at the upper band edge (just above), field peaks in low-index layers. Modulating a given layer thus preferentially couples to the band edge where field concentrates: high-index modulation excites \(f_0-f_\text{m}\), low-index modulation excites \(f_0+f_\text{m}\).

This intuitive Bloch-mode picture is quantitatively captured by the phase-matching condition for distributed parametric conversion across $  N  $ periods. For efficient coherent buildup of the sidebands, the accumulated phase mismatch per period should be an integer multiple of $  2\pi  $
\begin{equation}
	\Delta k_{\pm} \cdot \Lambda = 2\pi m, \quad m \in \mathbb{Z},
\end{equation}
where $  \Delta k_{\pm} = \beta_0 - \beta_{\pm}  $ is the wavevector mismatch between the carrier ($  \beta_0  $) and the generated sideband, with $  \beta_{\pm} \approx \beta_0 \mp \omega_{\rm m} / v_g  $. Here, $  v_g  $ is the effective group velocity of the Bloch mode at the Bragg frequency. In a quarter-wave Bragg stack, the standing-wave pattern of the carrier at $  \omega_0  $ exhibits a pronounced asymmetry: the electric field intensity $  |E|^2  $ is significantly higher in the low-index layers and suppressed in the high-index layers. This arises from the continuity of the displacement field at the interfaces combined with the $  \pi/2  $ phase shift per layer. Consequently, the effective group velocity $  v_{\rm g}  $ experienced by the carrier is a weighted average that is biased toward the faster velocity $  c/n_{\rm L}  $ (low-index layers), rather than the slower $ c/n_{\rm H}  $. This field asymmetry has important consequences for temporal phase matching. The product $  \Delta k_{\pm} \cdot \Lambda  $ therefore acquires slightly different magnitudes for up- and down-conversion
\begin{subequations}
\begin{equation}
	\Delta k_+ \cdot \Lambda \approx -\frac{\omega_\text{m}}{v_g} \Lambda \quad \text{(up-conversion)}, 
\end{equation}
\begin{equation}
	\Delta k_- \cdot \Lambda \approx +\frac{\omega_\text{m}}{v_g} \Lambda \quad \text{(down-conversion)}.
\end{equation}
\end{subequations}
\indent Since the carrier field is stronger in the low-index layers, temporal modulation applied to these layers experiences better spatial overlap with the pump field, leading to stronger coupling to the up-converted sideband ($+1 $) and more favorable phase-matching conditions for coherent accumulation. In contrast, modulation of the high-index layers (where the field is weaker) couples more effectively to the down-converted sideband ($-1$), with the phase accumulated across the period favoring constructive interference for that direction.
Thus, the layer-dependent conversion originates from two synergistic mechanisms: (i) the asymmetric Bloch-mode field distribution, which determines the local coupling strength to each sideband, and (ii) the resulting difference in temporal phase-matching conditions across the periodic structure. This combined effect explains why modulation of low-index layers selectively favors up-conversion, while modulation of high-index layers selectively favors down-conversion, a central result of this work that is confirmed by both the first-order analytical expressions and full transfer-matrix simulations.

\begin{figure}
	\begin{center}
\subfigure[]{\label{Fig:scatter_H}
		\includegraphics[width=0.8\columnwidth]{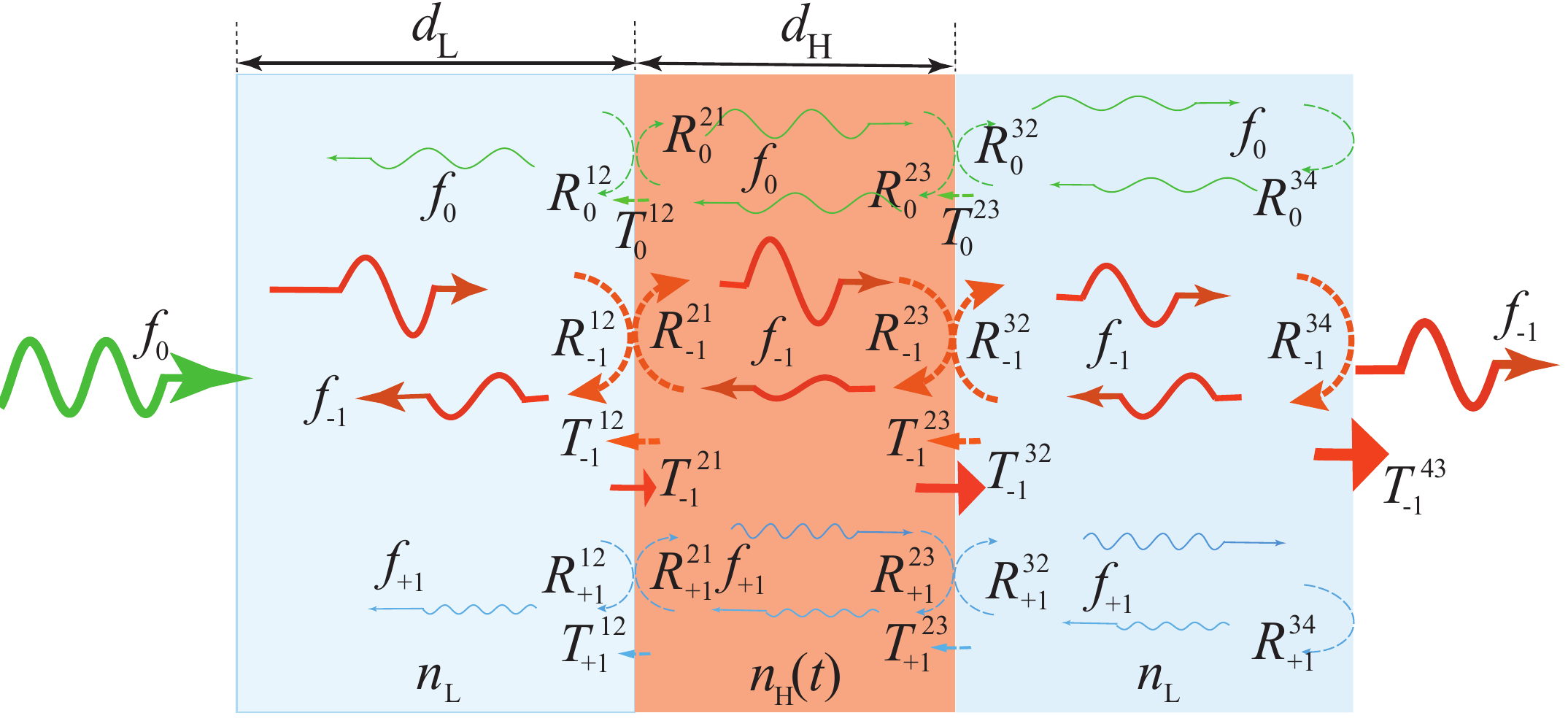}}
\subfigure[]{\label{Fig:scatter_L}
		\includegraphics[width=0.8\columnwidth]{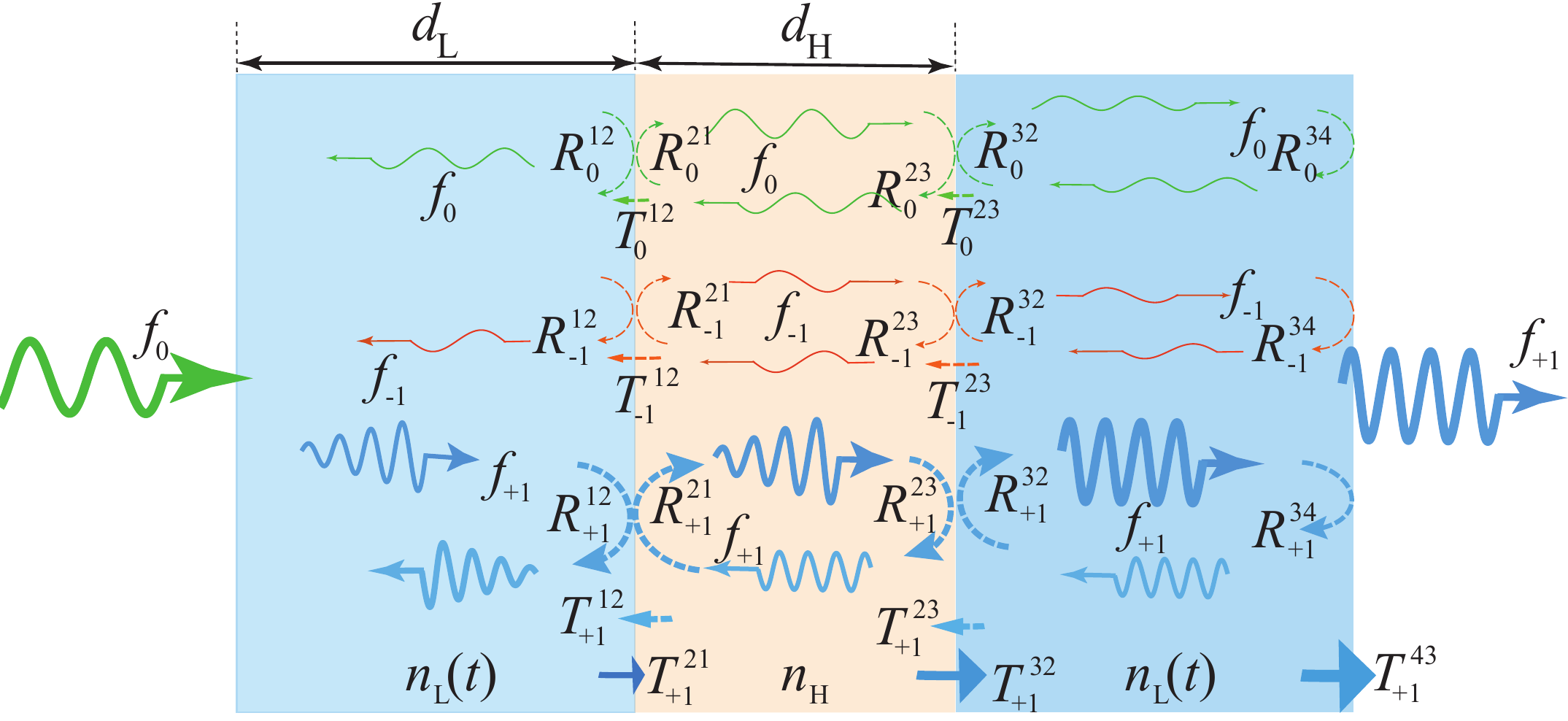}}
		\caption{Multiple-scattering diagram of a temporal Bragg grating. (a)~Down-conversion regime, where only the high-index layers are time-modulated. (b)~Up-conversion regime, where only the low-index layers are time-modulated.}
		\label{Fig:scatter}
	\end{center}
\end{figure}

Figures~\ref{Fig:scatter_H} presents space-time diagram of scattering in a Bragg grating with modulated high-index layers and static low-index layers, The lower band-edge mode concentrates field in high-index layers, preferentially generating \(f_{-1}\), which lies outside the stopband and propagates bidirectionally, while \(f_{+1}\) is trapped. Thus, a unified principle emerges: \textit{the modulated layer selects which sideband is generated; the stopband determines which escapes}. Bloch mode asymmetry ensures layer selectivity, while stopband filtering enables pure, bidirectional conversion—confirmed by full-wave simulations. Figure~\ref{Fig:scatter_L} presents space-time diagram of scattering in a Bragg grating with modulated low-index layers and static high-index layers, where the up-converted sideband \(f_{+1}\) dominates both output ports. This arises from two factors: (i) the Bloch mode at the upper band edge concentrates field in low-index layers, preferentially generating \(f_{+1}\); and (ii) \(f_{+1}\) lies outside the stopband, propagating freely in both directions (bold arrows). Conversely, \(f_{-1}\) lies inside the stopband, remains trapped, and the carrier \(f_0\) is strongly reflected—absent from both ports.

The refractive index of each layer may be static or time-modulated. We expand the electric field in a Floquet series
\begin{equation}
	E(z,t) = \sum_{n=-\infty}^{\infty} E_n(z) \, e^{-i \omega_n t},
	\label{eq:harmonic_expansion}
\end{equation}
where \(\omega_n = \omega_0 + n \omega_{\rm m}\) and \(E_n(z)\) denotes the full amplitude of the \(n\)-th harmonic. For weak modulation, only the fundamental (\(n=0\)) and first-order sidebands (\(n = \pm 1\)) carry significant power, so we truncate the expansion to these three harmonics. We define the slowly varying envelope \(\mathcal{E}_n(z)\) of each harmonic via
\begin{equation}
	E_n(z) = \mathcal{E}_n(z) \, \exp(i k_{j,n} z),
	\label{eq:sve_def}
\end{equation}
where \(k_{j,n} = \omega_n n_j / c\) is the propagation constant in layer \(j\) (\(n_j = n_{\rm L}\) or \(n_{\rm H}\)).

Within a time-modulated layer, the envelopes satisfy the coupled-mode equations obtained under the slowly varying envelope approximation
\begin{subequations}
	\begin{equation}
		\frac{d \mathcal{E}_n}{dz} = -i \kappa_j \left( e^{i\varphi_j} \mathcal{E}_{n+1} + e^{-i\varphi_j} \mathcal{E}_{n-1} \right),
		\label{eq:coupled_mode}
	\end{equation}
	with the coupling strength
	\begin{equation}
		\kappa_j = \frac{\omega_0 \Delta n_j}{2c}.
		\label{eq:coupling_strength}
	\end{equation}
\end{subequations}

The first term on the right-hand side of the full propagation (including the phase factor \(\exp(i k_{j,n} z)\)) accounts for natural phase accumulation, while the coupling term describes energy exchange between adjacent harmonics due to the temporal modulation. The modulation phase \(\varphi_j\) directly imprints on the generated sidebands. For the three-harmonic truncation (\(n = -1, 0, +1\)), the system can be written in matrix form for the slowly varying vector \(\boldsymbol{\mathcal{E}} = [\mathcal{E}_{-1}, \mathcal{E}_0, \mathcal{E}_{+1}]^T\)
\begin{equation}
	\frac{d \boldsymbol{\mathcal{E}}}{dz} = -i \mathbf{M}_j \boldsymbol{\mathcal{E}},
	\label{eq:matrix_form}
\end{equation}
where the coupling matrix is
\begin{equation}
	\mathbf{M}_j =
	\begin{bmatrix}
		0 & \kappa_j e^{-i\varphi_j} & 0 \\
		\kappa_j e^{i\varphi_j} & 0 & \kappa_j e^{-i\varphi_j} \\
		0 & \kappa_j e^{i\varphi_j} & 0
	\end{bmatrix}.
	\label{eq:coupling_matrix}
\end{equation}

The full propagation matrix through a layer of thickness \(d_j\) (including deterministic phase accumulation) reads
\begin{equation}
	\mathbf{P}_j = \mathbf{D}_j \, \exp(-i \mathbf{M}_j d_j),
	\label{eq:propagation_matrix}
\end{equation}
with the diagonal phase matrix
\begin{equation}
	\mathbf{D}_j = \operatorname{diag}\bigl( e^{i k_{j,-1} d_j}, \, e^{i k_{j,0} d_j}, \, e^{i k_{j,+1} d_j} \bigr),
\end{equation}
where for a static layer (\(\kappa_j = 0\)), \(\mathbf{M}_j = \mathbf{0}\) and \(\mathbf{P}_j = \mathbf{D}_j\).

At interfaces between layers \(i\) and \(j\), the refractive index changes discontinuously while the temporal modulation does not generate new frequencies. The boundary conditions are therefore diagonal in frequency space, with standard Fresnel coefficients evaluated at each harmonic frequency \(\omega_n\)
\begin{align}
	R_n^{ij} &= \frac{n_i - n_j}{n_i + n_j}, &
	T_n^{ij} &= \frac{2 n_i}{n_i + n_j},
	\label{eq:fresnel}
\end{align}
\begin{equation}
	\mathbf{T}^{ij} = \operatorname{diag}\bigl( T_{-1}^{ij}, \, T_0^{ij}, \, T_{+1}^{ij} \bigr).
	\label{eq:interface_matrix}
\end{equation}
\indent A unit cell of the temporal Bragg grating consists of one high-index and one low-index layer. The unit-cell transfer matrix (from left to right) is
\begin{equation}
	\mathbf{M} = \mathbf{T}^{\text{HL}} \mathbf{P}_\text{H} \mathbf{T}^{\text{LH}} \mathbf{P}_\text{L}.
	\label{eq:unit_cell}
\end{equation}
\indent Depending on which layers are modulated
\begin{itemize}
	\item Low-index layers modulated: \(\mathbf{P}_\text{L} = \mathbf{D}_\text{L} \exp(-i \mathbf{M}_\text{L} d_\text{L})\) (nondiagonal), \(\mathbf{P}_\text{H} = \mathbf{D}_\text{H}\) (diagonal).
	\item High-index layers modulated: \(\mathbf{P}_\text{H} = \mathbf{D}_\text{H} \exp(-i \mathbf{M}_\text{H} d_\text{H})\) (nondiagonal), \(\mathbf{P}_\text{L} = \mathbf{D}_\text{L}\) (diagonal).
\end{itemize}
and for a grating with \(N\) periods, the total transfer matrix is
\begin{equation}
	\mathbf{M}_{\rm tot} = \mathbf{M}^N.
	\label{eq:total_matrix}
\end{equation}
\indent The output harmonic amplitudes for an incident carrier wave of amplitude \(A_0^{\rm inc}\) (with zero incoming sidebands) are
\begin{equation}
	\begin{bmatrix}
		E_{-1}^{\rm out} \\ E_0^{\rm out} \\ E_{+1}^{\rm out}
	\end{bmatrix}
	= \mathbf{M}_{\rm tot}
	\begin{bmatrix}
		0 \\ A_0^{\rm inc} \\ 0
	\end{bmatrix},
	\label{eq:harmonic_output}
\end{equation}
and the quarter-wave condition for the carrier (\(n=0\)) in each layer, that is,
\begin{subequations}
	\begin{align}
		k_{\rm L,0} d_{\rm L} &= \frac{\omega_0 n_{\rm L} d_{\rm L}}{c} = \frac{\pi}{2}, \\
		k_{\rm H,0} d_{\rm H} &= \frac{\omega_0 n_{\rm H} d_{\rm H}}{c} = \frac{\pi}{2},
	\end{align}
	\label{eq:quarter_wave_phase}
\end{subequations}
produces the Bragg stopband for the fundamental frequency.For weak modulation in the low-index layers (\(\kappa_{\rm L} d_{\rm L} \ll 1\)) and in the high-index layers (\(\kappa_{\rm H} d_{\rm H} \ll 1\)), first-order perturbation theory yields simple closed-form expressions for the generated sidebands
\begin{subequations} 
	\begin{align}
		T_{+1} &\approx -i N \kappa_{\rm L} d_{\rm L} \, T_0^{\rm Bragg} \, e^{i \phi_{\rm L}},  \label{eq:first_order_up_T} \\
		R_{+1} &\approx -i N \kappa_{\rm L} d_{\rm L} \, R_0^{\rm Bragg} \, e^{i \phi_{\rm L}},  \label{eq:first_order_up_R} \\
		T_{-1} &\approx -i N \kappa_{\rm H} d_{\rm H} \, T_0^{\rm Bragg} \, e^{-i \phi_{\rm H}}, \label{eq:first_order_down_T} \\
		R_{-1} &\approx -i N \kappa_{\rm H} d_{\rm H} \, R_0^{\rm Bragg} \, e^{-i \phi_{\rm H}}. \label{eq:first_order_down_R},
	\end{align}
\end{subequations}
\noindent where \(T_0^{\rm Bragg}\) and \(R_0^{\rm Bragg}\) are the carrier transmission and reflection coefficients of the unmodulated Bragg grating. For a high-contrast grating with many periods, \(T_0^{\rm Bragg} \approx 0\) and \(R_0^{\rm Bragg} \approx 1\). When a period-to-period phase gradient \(\Delta\phi\) is introduced across successive periods, the phase in the \(p\)-th period is $\phi_p = \phi_{\rm H} + p\,\Delta\phi~\text{(for high-index modulation)}~\text{or}~\phi_p = \phi_{\rm L} + p\,\Delta\phi ~\text{(for low-index modulation)}$. Then, the sideband transmission becomes
\begin{subequations}
\begin{equation}
	T_{\pm}(\Delta\phi) \approx \gamma A_0^{\rm inc} e^{i\phi_0} \sum_{p=0}^{N-1} (\tau_0 e^{i\Delta\phi})^p (\tau_{\pm})^{N-1-p},
	\label{eq:phase_gradient}
\end{equation}
which evaluates to the geometric series
\begin{equation}
	T_{\pm}(\Delta\phi) = \gamma A_0^{\rm inc} e^{i\phi_0} \frac{ (\tau_{\pm})^N - (\tau_0 e^{i\Delta\phi})^N }{ \tau_{\pm} - \tau_0 e^{i\Delta\phi} }.
\end{equation}
\end{subequations}
\indent Resonant enhancement occurs when the temporal phase-matching condition
\begin{equation}
	\Delta\phi = \arg\left( \frac{\tau_{\pm}}{\tau_0} \right),
	\label{eq:phase_matching}
\end{equation}
is satisfied. This transforms the temporal Bragg grating into a temporally phased array, enabling electronic control of output amplitude, switching, and spectral shaping.

\section{Results}

\begin{figure}
	\begin{center}
		\includegraphics[width=1\columnwidth]{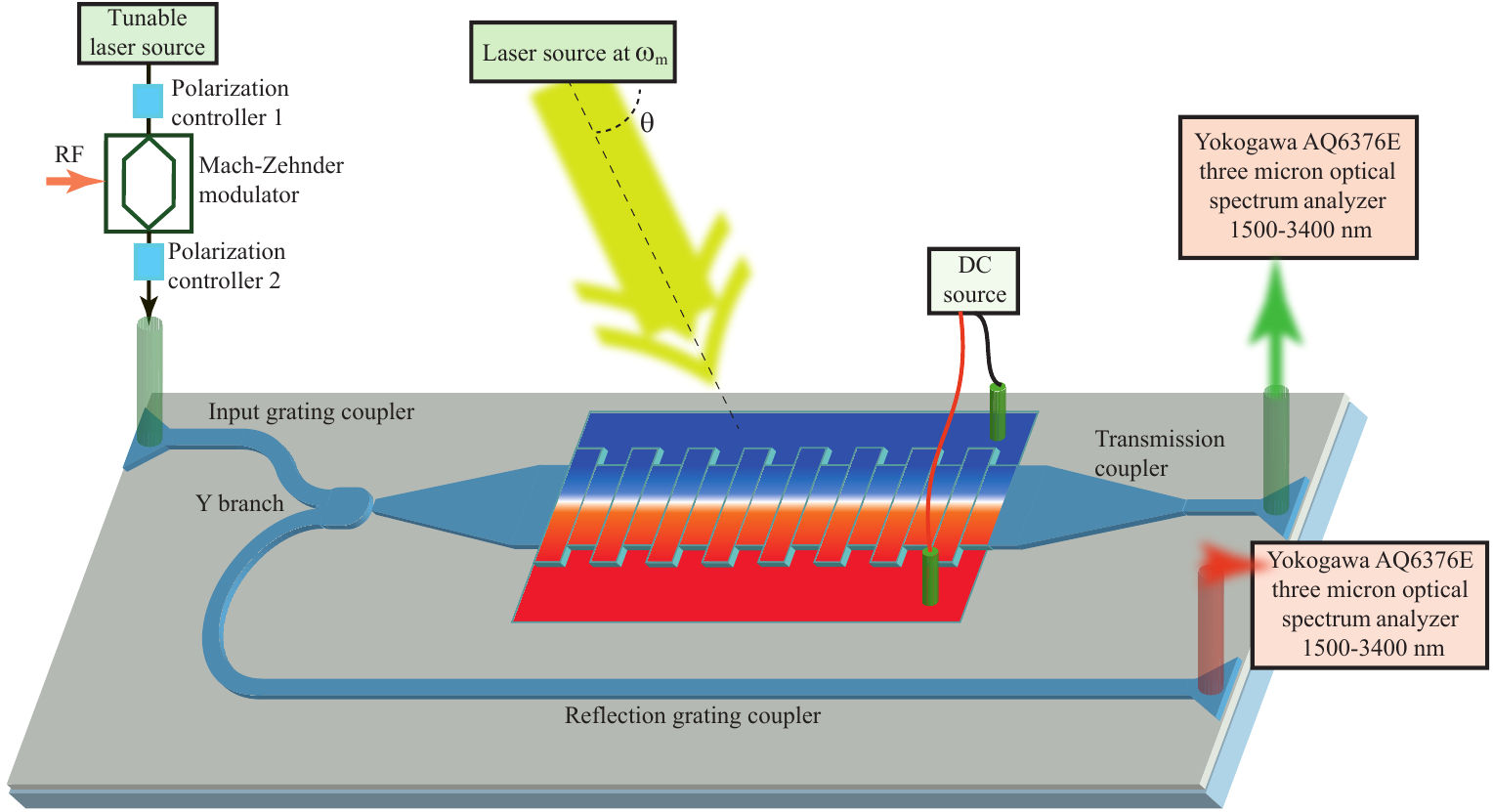}
		\caption{Schematic of the experimental setup for the temporal Bragg grating frequency converter in Fig.~\ref{fig:fab}. The input signal at $\omega_0 $ is prepared using a tunable laser source, polarization controllers, and Mach-Zehnder modulator. The modulation pump is incident at angle $\theta$ on the device. DC bias is applied to the PN junctions. Transmitted and reflected spectra are measured using a mid-infrared optical spectrum analyzer.}
		\label{Fig:setup}
	\end{center}
\end{figure}

An experimental setup to characterize the proposed temporal Bragg grating frequency converter is shown in Fig.~\ref{Fig:setup}. A tunable laser source generates the continuous-wave input signal at frequency $\omega_0 \approx 150$ THz. The signal passes through polarization controllers (PC1 and PC2) and a Mach-Zehnder modulator (MZM) driven by an RF input before being coupled into the device via an input grating coupler. The modulation pump beam at $\omega_{\rm m} = 30$ THz is generated separately and directed onto the grating at a controlled oblique angle $\theta$ to simultaneously induce temporal refractive index modulation in the doped sections and the desired phase gradient $\Delta\phi$ along the structure. A DC source provides bias voltage to the lateral PN junctions in the modulated sections. Both transmitted and reflected signals are collected through their respective grating couplers and analyzed using a mid-infrared optical spectrum analyzer (Yokogawa AQ6376E, covering 1500–3400 nm) to resolve the carrier frequency and the generated sidebands at $\omega_0 \pm \omega_{\rm m}$.

\begin{figure}
	\begin{center}
		\includegraphics[width=0.6\columnwidth]{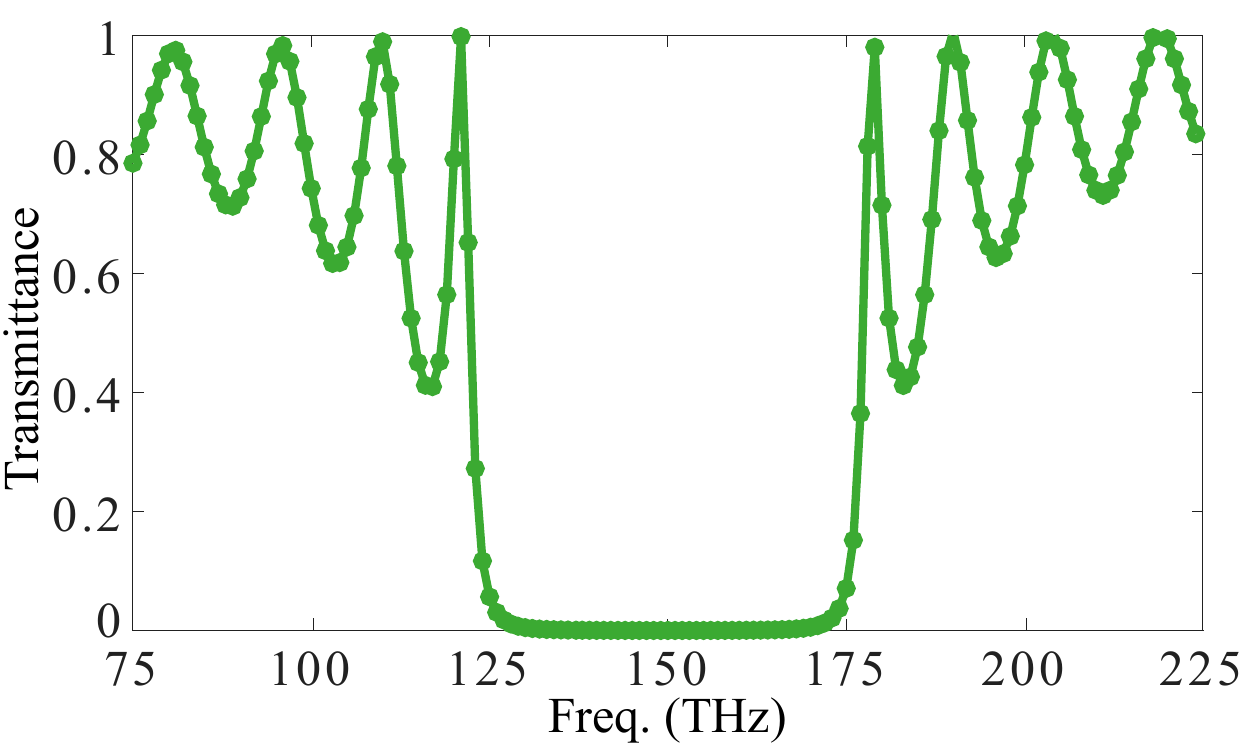}
		\caption{Full-wave simulation results for transmittance as a function of frequency for the passive (unmodulated) Bragg grating consisting of \(N = 8\) periods of alternating high-index (\(n_{\rm H} = 2.5\)) and low-index (\(n_{\rm L} = 1.5\)) layers, each satisfying the quarter-wave condition at the design frequency \(f_0 = 150\) THz.}
		\label{Fig:Bragg}
	\end{center}
\end{figure}

\begin{figure}
	\begin{center}
		\subfigure[]{\label{Fig:DCa}
			\includegraphics[width=0.3\columnwidth]{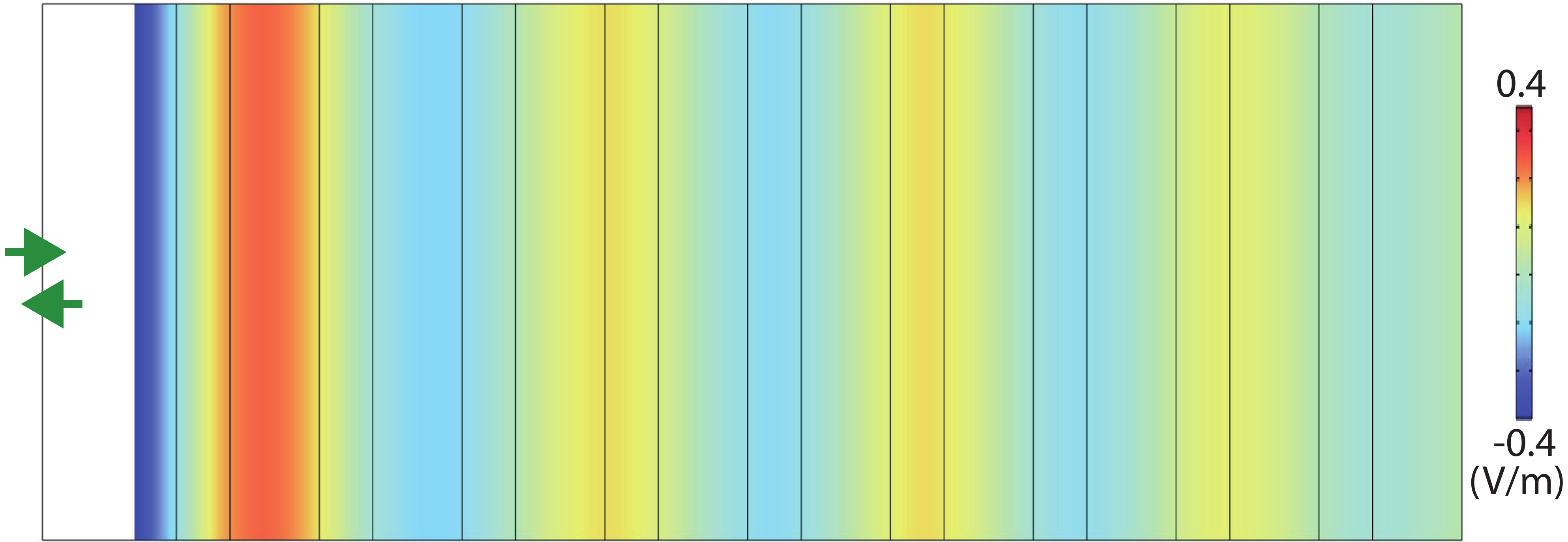}}
			\subfigure[]{\label{Fig:DCb}
		\includegraphics[width=0.3\columnwidth]{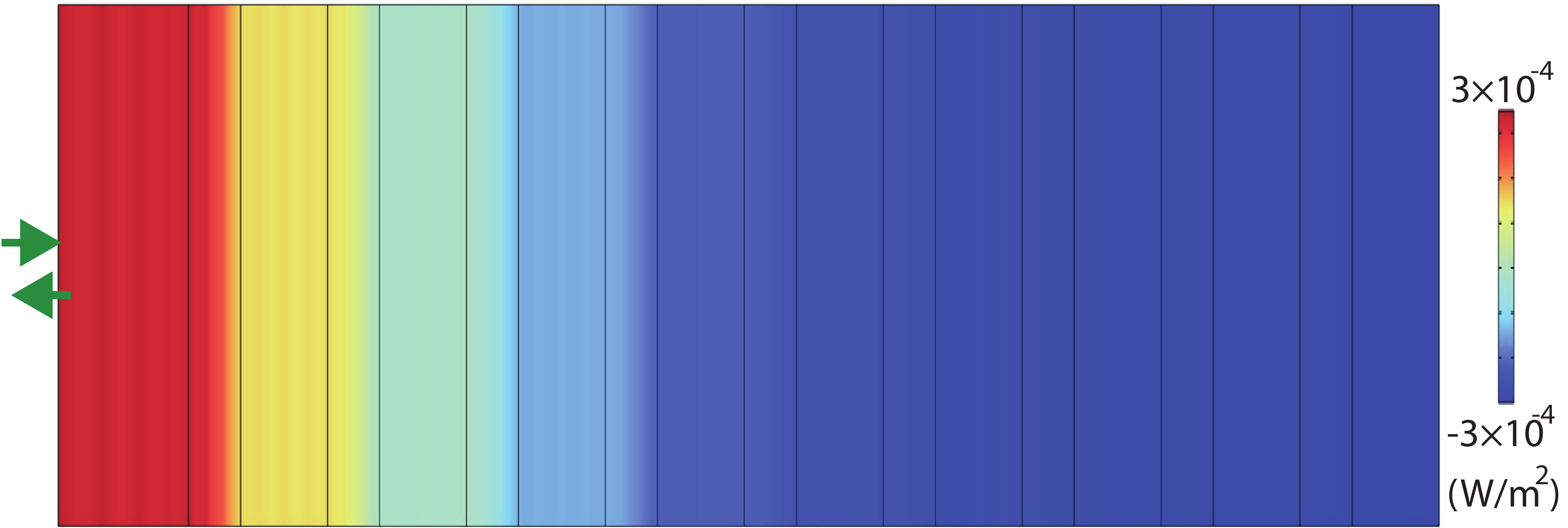}}
			\subfigure[]{\label{Fig:DCc}
	\includegraphics[width=0.3\columnwidth]{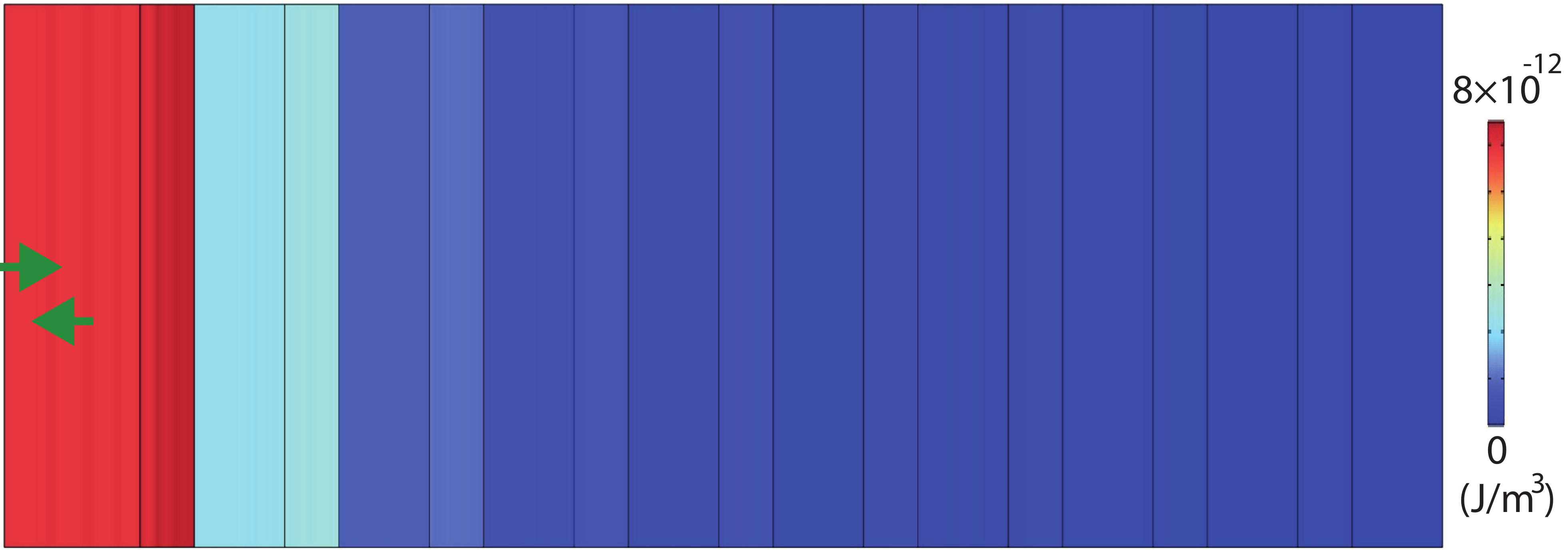}}
		\subfigure[]{\label{Fig:DCd}
			\includegraphics[width=0.3\columnwidth]{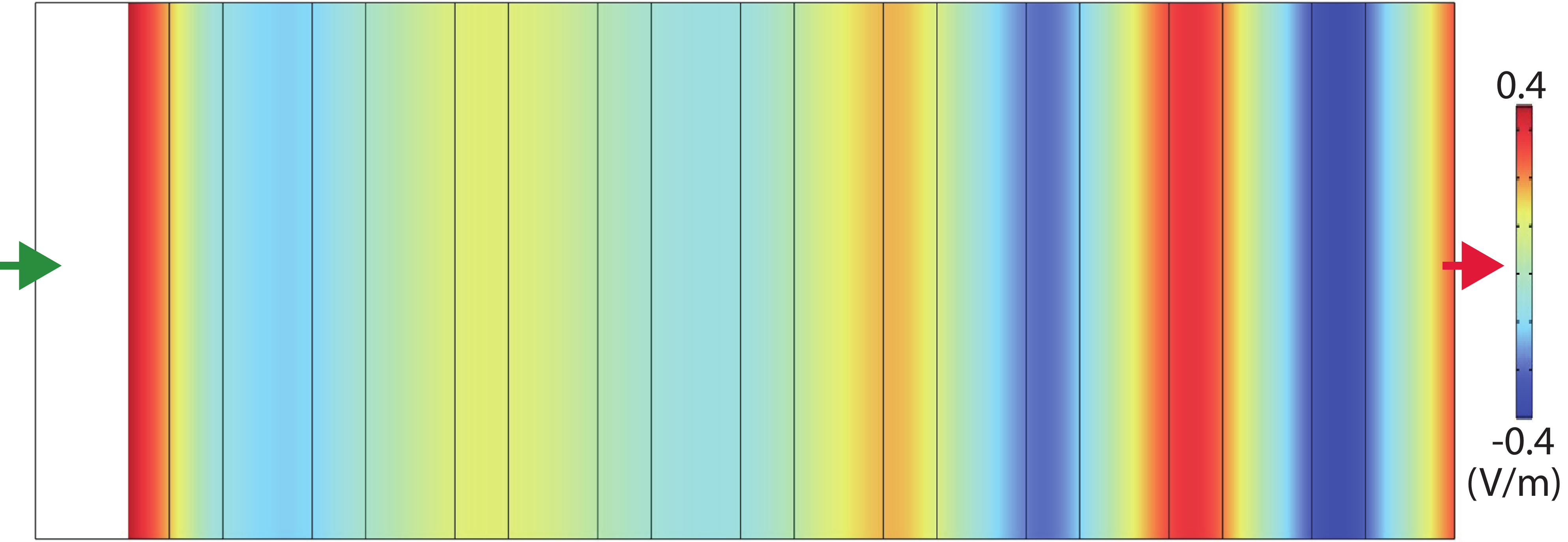}}
			\subfigure[]{\label{Fig:DCe}
		\includegraphics[width=0.3\columnwidth]{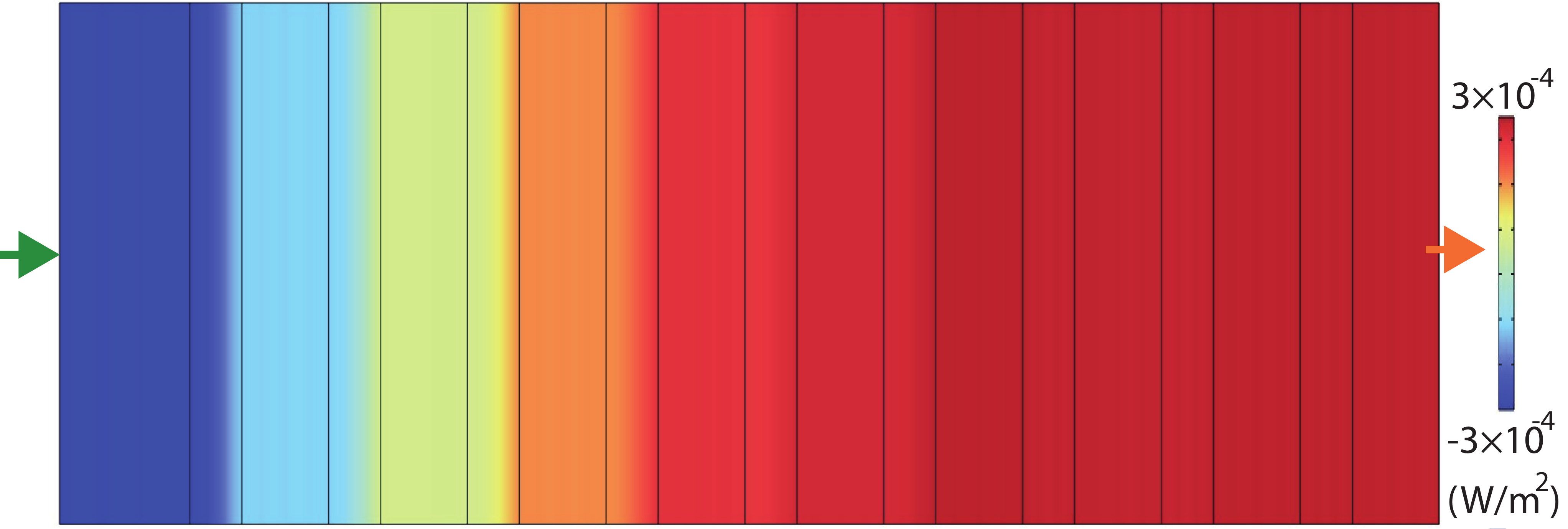}}
				\subfigure[]{\label{Fig:DCf}
		\includegraphics[width=0.3\columnwidth]{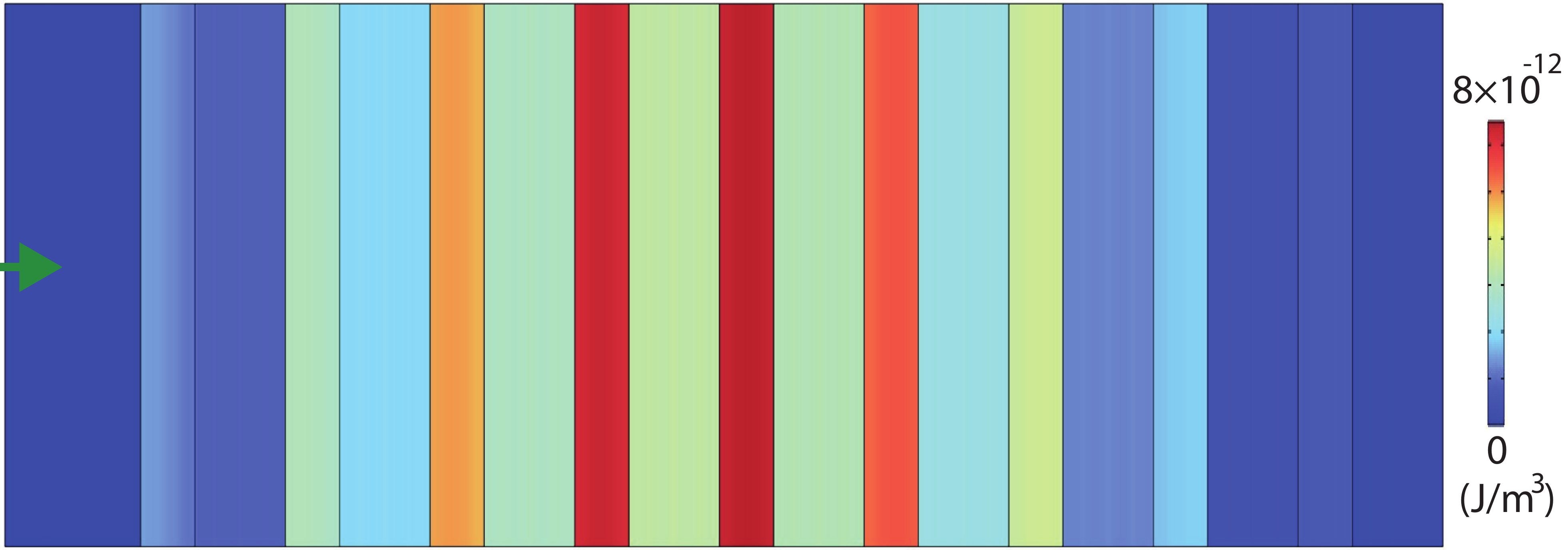}}
				\subfigure[]{\label{Fig:DCg}
	\includegraphics[width=0.48\columnwidth]{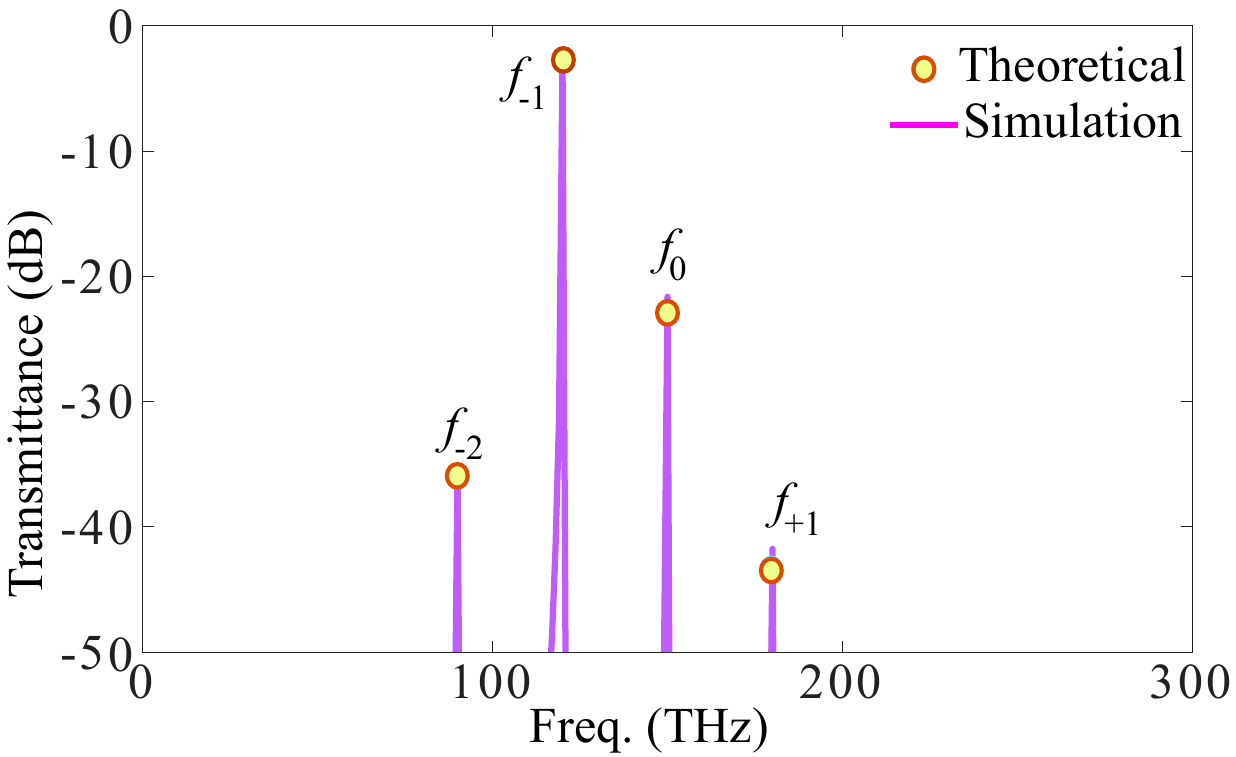}}
				\subfigure[]{\label{Fig:DCh}
	\includegraphics[width=0.48\columnwidth]{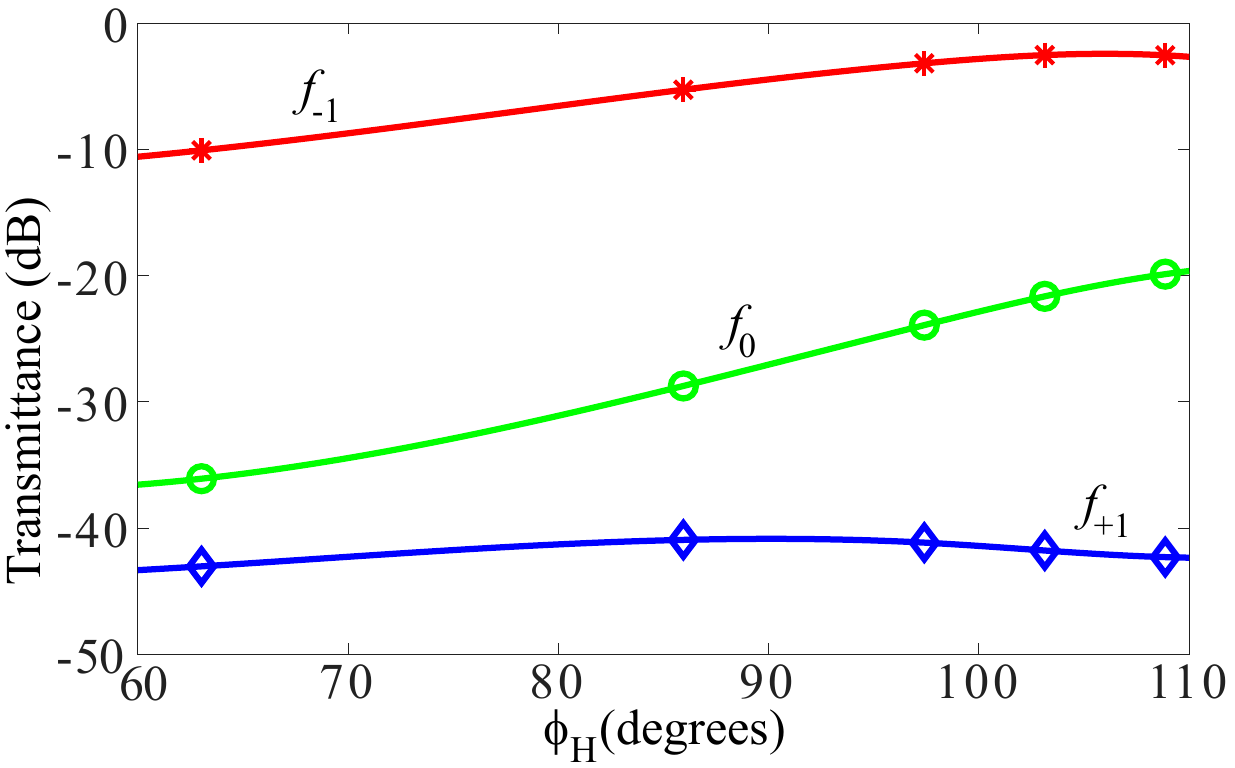}}
\caption{Full-wave simulation results for down-conversion mode (high-index layers modulated), with $N=8$, $n_\text{H}(t) = 2.5[1 + 0.12 \cos(0.2\omega_0 t - 103.2^\circ)]$, and $n_\text{L}=1.5$. (a)~Electric field \(E_z\) at the incident frequency \(f_0 = 150\) THz. (b)~Time-averaged power flow at \(f_0 \). (c)~Time-averaged energy density at \(f_0 \). (d)~Electric field \(E_z\) at \(f_{-1} =  120\) THz. (e)~Time-averaged power flow at \(f_{-1}\). (f)~Time-averaged energy density at \(f_{-1}\). (g)~Output spectrum, comparing theoretical results from the coupled-mode model with full-wave simulation results. (h)~Transmittance versus modulation phase \(\varphi_\text{H}\) at $f_{0}, f_{-1}$ and $f_{+1}$.}
		\label{Fig:DC}
	\end{center}
\end{figure}

We first characterize the passive Bragg grating. Figure~\ref{Fig:Bragg} shows its transmittance spectrum (\(N=8\) periods). A deep rejection notch at \(f_0=150\) THz (transmittance \(<10^{-3}\), \(<-30\) dB) confirms that the carrier lies at the center of a photonic stopband. This spectral selectivity is key, where the carrier is blocked while sidebands \(f_{\pm1}\) transmit freely. Outside the stopband, Fabry-Perot resonances from the finite-length grating produce transmission peaks approaching unity at frequencies where an integer number of half-wavelengths fit within \(L=N\Lambda\).

To further analyze the structure, we perform full-wave finite-difference time-domain (FDTD) simulations of the temporal Bragg grating. The Bragg structure consists of \(N = 8\) periods, comprising $8$ high-index and $8$ low-index layers, with a center frequency of \(f_0 = 150\) THz corresponding to a Bragg wavelength \(\lambda_\text{B} = 2\;\mu\text{m}\) at the center of the stopband. The layers are time-modulated according to \(n_\text{L} = 1.5\) (static) and $n_\text{H}(t) = 2.5[1 + 0.12 \cos(0.2\omega_0 t - 103.2^\circ)]$, where the modulation frequency \(\omega_\text{m} = 0.2\omega_0 = 0.4\pi \times 150\) THz (i.e., \(f_\text{m} = 0.2 f_0 = 30\) THz) targets down-conversion to \(f_{-1} \). 

At the incident frequency \(f_0=150\) THz (Figs.~\ref{Fig:DCa}–\ref{Fig:DCc}), the electric field forms a standing-wave pattern on the left with exponential decay into the grating—clear signatures of Bragg reflection. Power flow is confined to the left side (red positive, blue negative), confirming zero transmission. Energy density vanishes on the right, establishing that the carrier is completely blocked by the stopband. At the down-converted frequency \(f_{-1}\) (Figs.~\ref{Fig:DCd}–\ref{Fig:DCf}), the behavior reverses dramatically. Strong field appears on both sides (Fig.~\ref{Fig:DCd}), indicating bidirectional generation. Power flows positively to the right and negatively to the left (Fig.~\ref{Fig:DCe}), confirming equal forward and backward conversion. Energy density (Fig.~\ref{Fig:DCf}) peaks in the high-index layers near the grating center, revealing resonant buildup: the down-converted wave experiences cavity-like enhancement, with forward and backward components interfering constructively in the middle periods.

Figure~\ref{Fig:DCg} presents the output spectrum of the temporal Bragg grating frequency down-converter. Excellent agreement is observed between the theoretical results from the coupled-mode model and the full-wave simulation results for both the carrier and the generated sidebands. This demonstrates the accuracy of the developed analytical framework. The output spectrum is dominated by a strong peak at \(f_{-1}\), with negligible power at \(f_0\) and \(f_{+1}\), confirming a pure and efficient down-conversion. The absence of \(f_{+1}\) validates layer selectivity, where modulating high-index layers preferentially couples to the lower band edge, generating \(f_{-1}\) while suppressing \(f_{+1}\). Figure~\ref{Fig:DCh} demonstrates phase control of conversion efficiency. As \(\varphi_\text{H}\) varies from \(60^\circ\) to \(110^\circ\), the down-converted power \(f_{-1}\) tunes sinusoidally by \(7.48\) dB, peaking at \(\varphi_\text{H}\approx103.2^\circ\). The carrier \(f_0\) remains suppressed below \(-20\) dB across all phases, and \(f_{+1}\) stays below \(-42\) dB (\(>30\) dB below the peak), which confirms that the stopband blocks the carrier regardless of modulation, and that the up-converted sideband remains trapped with no resonant enhancement.

\begin{figure}
	\begin{center}
		\subfigure[]{\label{Fig:UCa}
			\includegraphics[width=0.3\columnwidth]{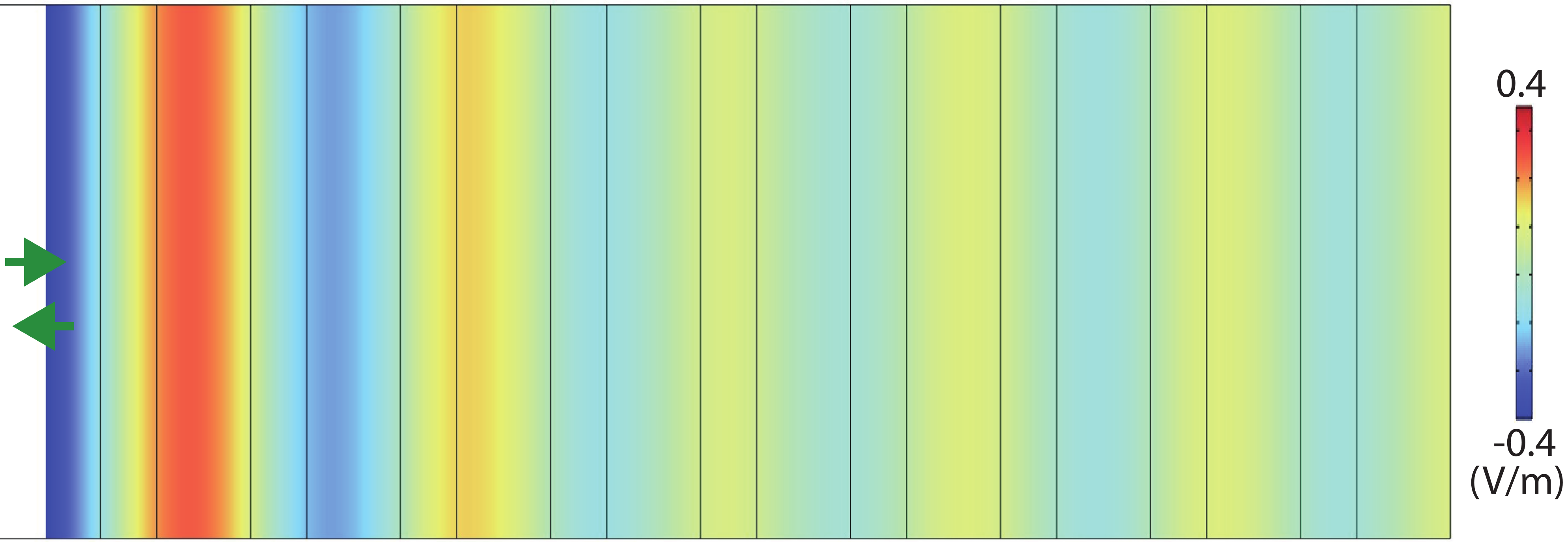}}
		\subfigure[]{\label{Fig:UCb}
			\includegraphics[width=0.3\columnwidth]{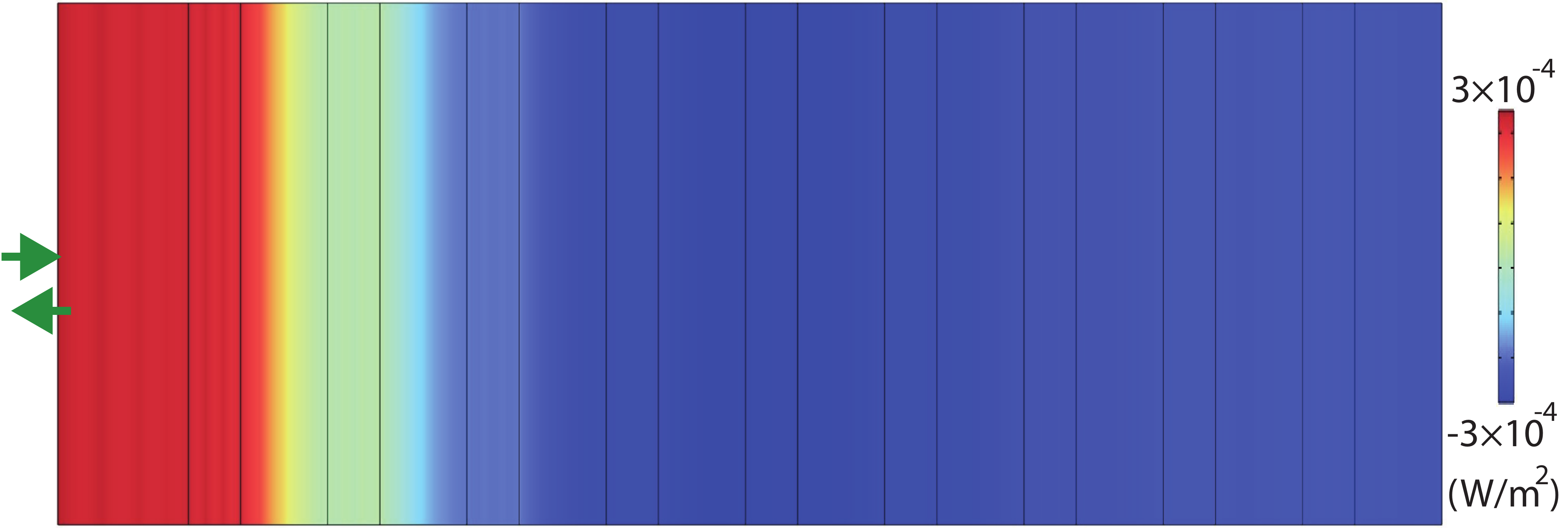}}
		\subfigure[]{\label{Fig:UCc}
			\includegraphics[width=0.3\columnwidth]{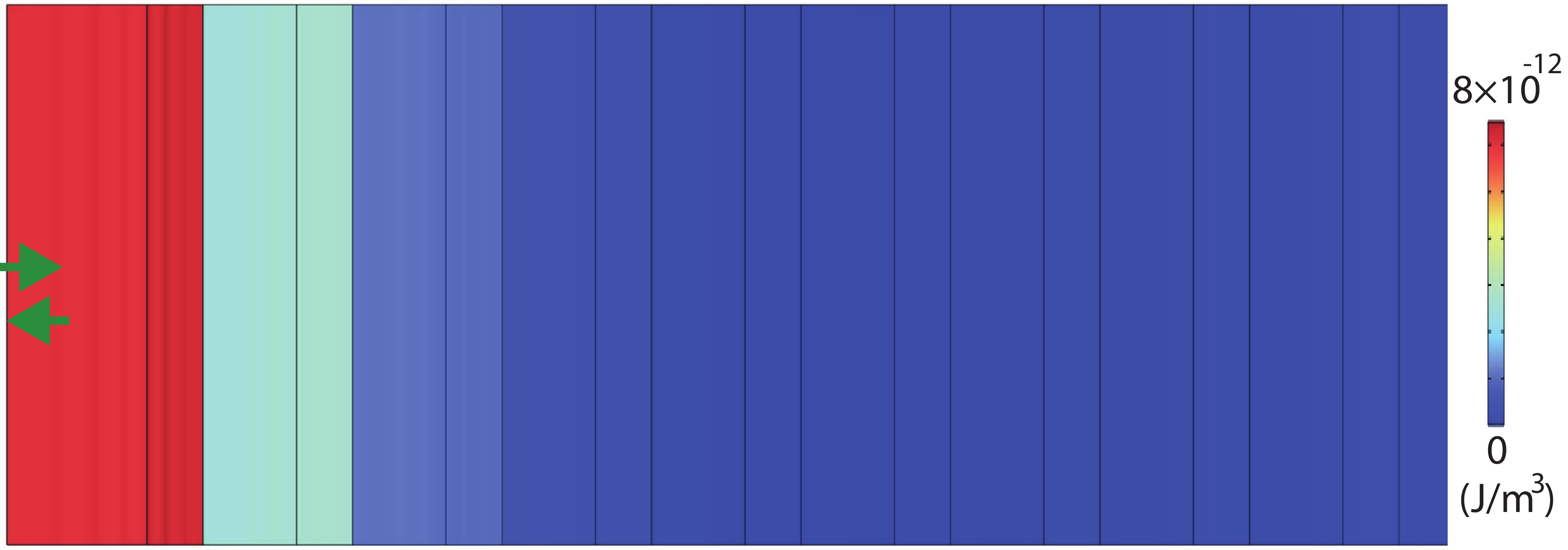}}
		\subfigure[]{\label{Fig:UCd}
			\includegraphics[width=0.3\columnwidth]{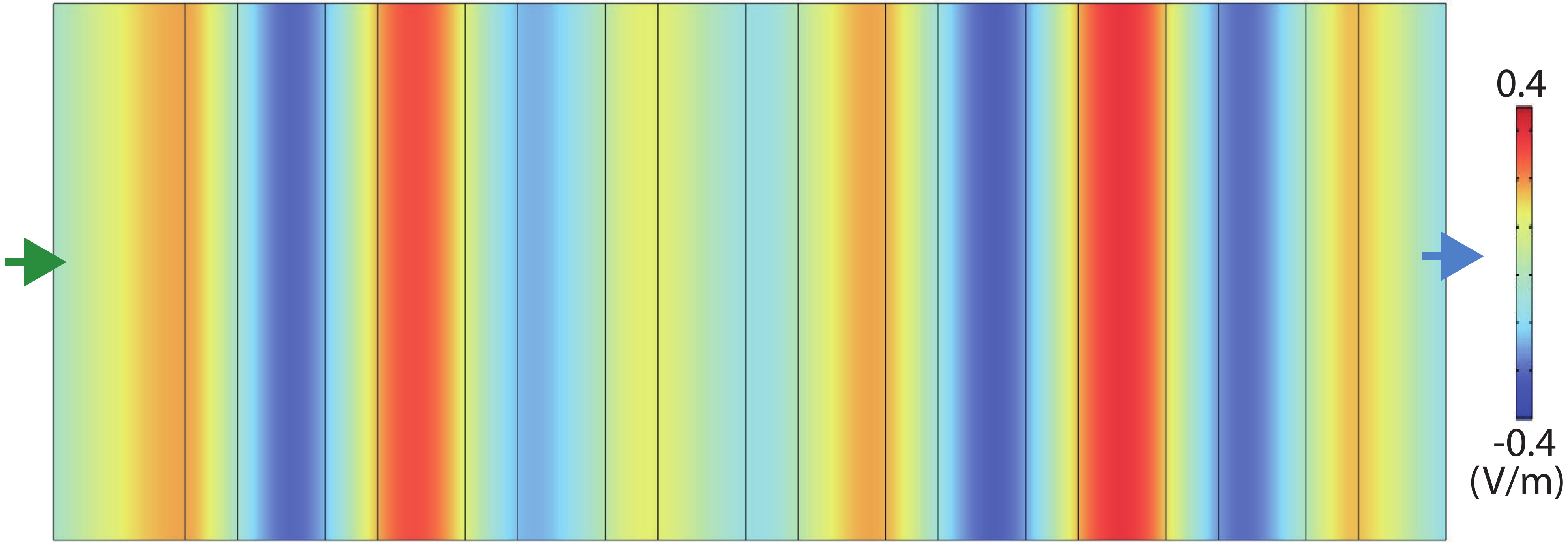}}
		\subfigure[]{\label{Fig:UCe}
			\includegraphics[width=0.3\columnwidth]{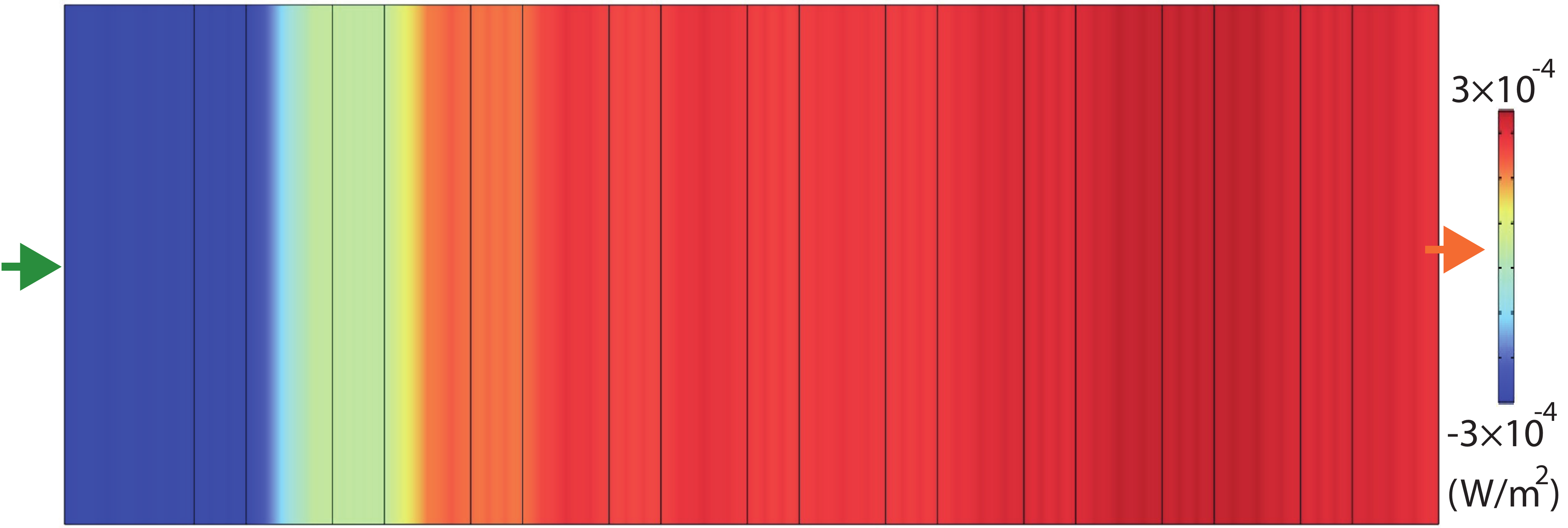}}
		\subfigure[]{\label{Fig:UCf}
			\includegraphics[width=0.3\columnwidth]{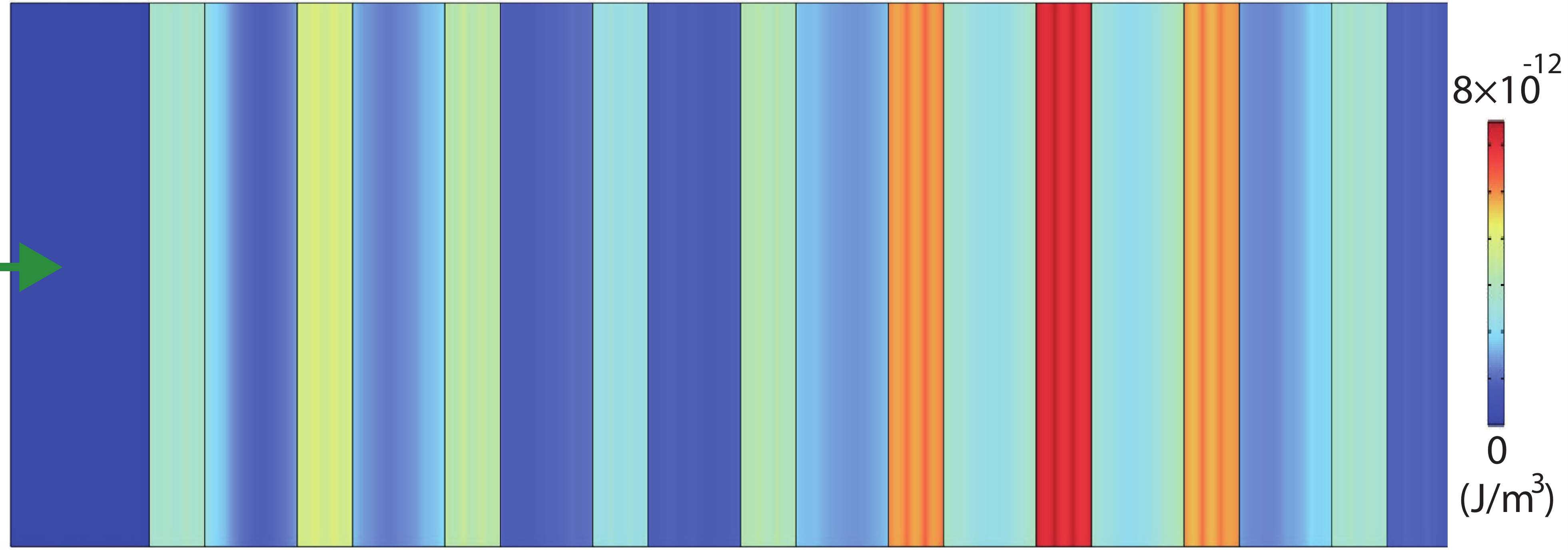}}
		\subfigure[]{\label{Fig:UCg}
			\includegraphics[width=0.48\columnwidth]{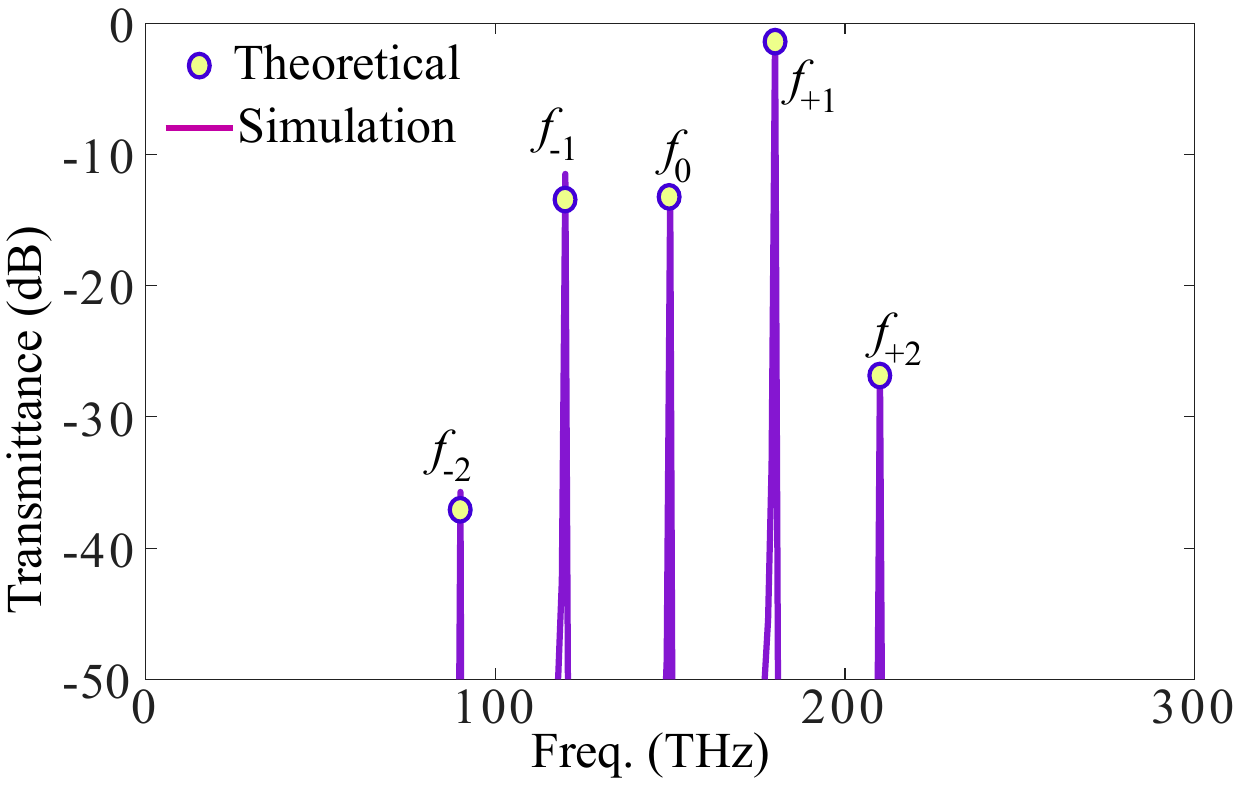}}
		\subfigure[]{\label{Fig:UCh}
			\includegraphics[width=0.48\columnwidth]{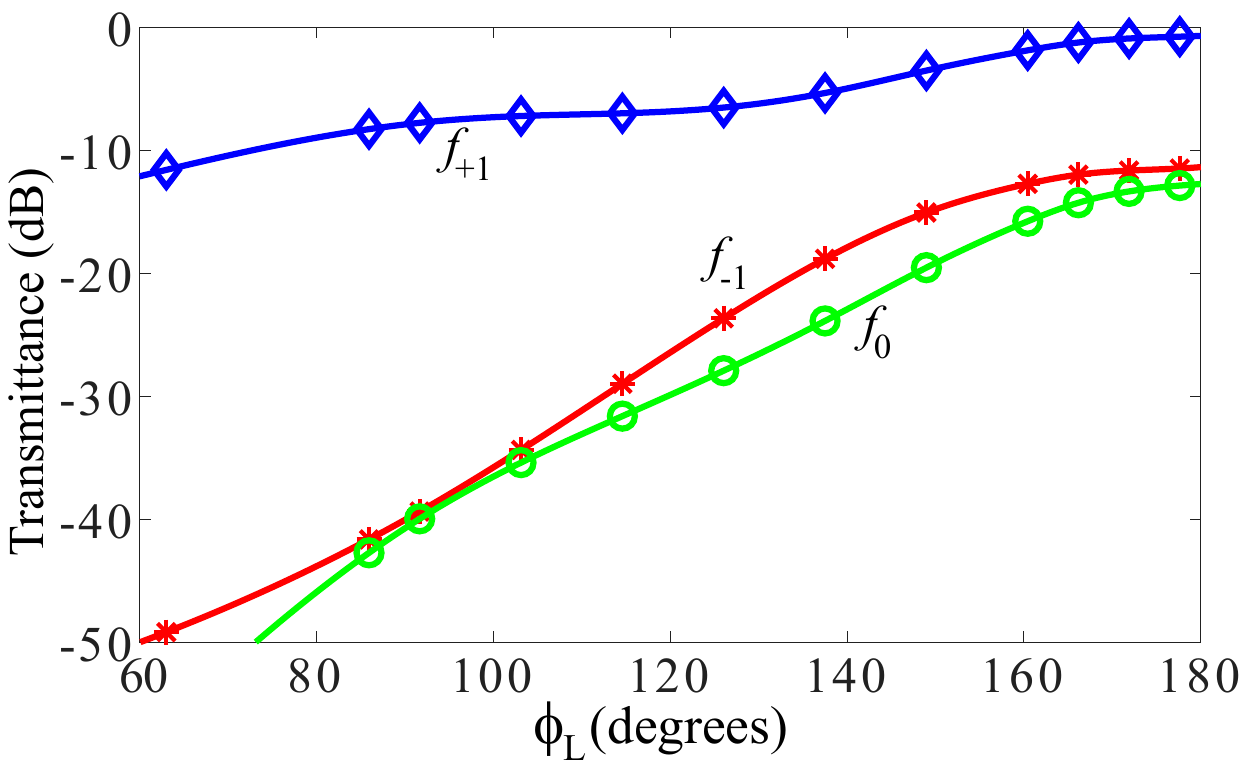}}
		\caption{Full-wave simulation results for up-conversion mode, with $n_{\rm L}(t) = 1.5[1 + 0.12 \cos(0.2\,\omega_0 t + \varphi_{\rm L})$, $n_{\rm H}=2.5$, and \(\varphi_{\rm L} = 166.2^\circ\), and high-index layers are static. The incident wave at frequency \(f_0 \) enters from the left. (a)~Electric field \(E_z\) at the incident frequency \(f_0 = 150\) THz. (b)~Time-averaged power flow at \(f_0 \), indicating power flows only on the left side. (c)~Time-averaged energy density at \(f_0\). Energy is concentrated on the left side, with an exponentially decaying envelope into the structure, characteristic of evanescent penetration into a stopband. (d) Electric field \(E_z\) at the up-converted frequency \(f_{+1} = 180\) THz. (e)~Time-averaged power flow at \(f_{+1} \). (f)~Time-averaged energy density at \(f_{+1} \). (g) Output spectrum, comparing theoretical results from the coupled-mode model with full-wave simulation results. (h)~Dependence of transmittance on modulation phase \(\varphi_{\rm L}\).}
		\label{Fig:UC}
	\end{center}
\end{figure}

We next present simulations for up-conversion, where low-index layers are time-modulated, where (\(n_{\rm L}(t)=1.5[1+0.12\cos(0.2\omega_0 t+\varphi_{\rm L})]\), $n_{\rm H}=2.5$, and \(\varphi_{\rm L}=166.2^\circ\)). Figure~\ref{Fig:UC} mirrors Fig.~\ref{Fig:DC} for direct comparison. At \(f_0\) (Figs.~\ref{Fig:UCa}–\ref{Fig:UCc}), the response is identical to the down-conversion case: standing-wave patterns, power confined to the left, and exponentially decaying energy density. This confirms that the carrier is blocked by the Bragg stopband regardless of which layer is modulated. At \(f_{+1}\) (Figs.~\ref{Fig:UCd}–\ref{Fig:UCf}), strong field appears on both sides (Fig.~\ref{Fig:UCd}), with power flowing positively rightward and negatively leftward (Fig.~\ref{Fig:UCe}), which shows bidirectional generation. The energy density (Fig.~\ref{Fig:UCf}) peaks in high-index layers near the output, increasing monotonically from left to right. This contrasts sharply with down-conversion, where energy peaks at the center. The asymmetry originates from the phase-mismatch sign: \(\Delta k_- > 0\) for down-conversion yields resonant center buildup; \(\Delta k_+ < 0\) for up-conversion gives monotonic growth toward the output. The output spectrum (Fig.~\ref{Fig:UCg}) is dominated by \(f_{+1}\), with \(f_0\) and \(f_{-1}\) suppressed by \(>11.6\) dB and \(>11\) dB, which represents a pure up-conversion. Figure~\ref{Fig:UCh} shows the output versus \(\varphi_\text{L}\): the up-converted power tunes sinusoidally by \(12\) dB, peaking at \(\varphi_\text{L}\approx166^\circ\). The carrier and \(f_{-1}\) remain suppressed below \(-11.6\) dB and \(-11\) dB across all phases, demonstrating selective, phase-independent up-conversion.

Such a layer-selective frequency conversion arises from the interplay between the spatial field distribution of the Bragg mode and the local temporal index modulation. In a quarter-wave Bragg stack at the carrier frequency \(\omega_0\), the electric field forms a standing-wave pattern with significantly higher intensity in the low-index (\(n_{\rm L}\)) layers compared to the high-index (\(n_{\rm H}\)) layers. This is a direct consequence of the continuity of the displacement field \(D = \varepsilon E\) across interfaces and the \(\pi/2\) phase shift per layer. When only the low-index layers are temporally modulated, the strong local electric field overlaps maximally with the modulation, enabling efficient parametric energy transfer from the carrier to the higher-frequency sideband (\(+1\), up-conversion). Conversely, modulation of the high-index layers occurs where the field intensity is lower, and the accumulated spatial phase across each period shifts the effective temporal phase-matching condition, favoring energy transfer to the lower-frequency sideband (\(-1\), down-conversion). The opposite signs in the phase factors \(e^{\pm i \phi_j}\) originate from the complex representation of the cosine modulation and the direction of frequency shift relative to the carrier. This behavior is robust and is accurately captured by the full transfer-matrix model beyond the first-order approximation, confirming that the selectivity is a fundamental feature of spatiotemporal Bragg structures rather than a perturbative artifact.

The proposed silicon rib waveguide implementation offers several advantages compared to existing frequency conversion technologies. Unlike conventional nonlinear frequency converters requiring phase-matched nonlinear crystals or lengthy waveguides, our device achieves frequency conversion via temporal modulation with modest modulation depths ($\delta= 0.12$), relaxing fabrication tolerances. Fabrication complexity is comparable to standard silicon photonic modulators: the device requires only two lithographic steps (waveguide definition and selective doping), avoiding more demanding processes such as epitaxial regrowth, wafer bonding, or the integration of nonlinear materials. Compared to lithium-niobate-based converters, the CMOS-compatible silicon platform offers lower cost and easier integration with electronic control circuits. The primary fabrication challenge is the precise control of doping profiles to achieve the desired index contrast and PN junction performance. However, this is within the capabilities of commercial foundry processes.

\section{Conclusion}
We introduced the Bragg Frequency Converter, a spatiotemporal grating that enables selective parametric frequency conversion. Our theoretical and numerical results show that modulation of the high-index layers produces efficient down-conversion, while modulation of the low-index layers yields up-conversion. The device reflects the carrier frequency while allowing the generated sidebands to propagate in both transmission and reflection directions. Full-wave simulations confirm the theoretical predictions and reveal distinct energy accumulation behaviors for up- and down-conversion. A practical implementation and an experimental measurement setup is proposed for future experimental validation. The proposed concept offers clean, spurious-free frequency conversion in a compact all-dielectric platform using modest modulation depths.

\bibliographystyle{IEEEtran}
\bibliography{Taravati_Reference.bib}

\vspace{-1cm}
\begin{IEEEbiography}[{\includegraphics[width=0.95in,height=1.2in,clip,keepaspectratio]{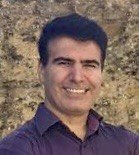}}]{Sajjad Taravati} (Senior Member, IEEE) is an Assistant Professor (UK Lecturer) at the University of Southampton, UK, where he leads a research group focused on dynamic metasurfaces for next-generation communication and quantum systems. Over the past decade, his research at the University of Toronto, Université de Montréal, Concordia University, the University of Oxford, and the University of Southampton has advanced nonreciprocal electromagnetics, spatiotemporal metasurfaces, and quantum computing, resulting in over 90 publications and patents. He serves as a Technical Committee Member of the IEEE Antennas and Propagation Society (TC-5: Electromagnetics and Fundamentals) and is the Technological Founder of LATYS Intelligence, a company commercializing dynamic metasurface technologies based on his patents.
\end{IEEEbiography}

\end{document}